\documentclass[journal,10pt,twocolumn]{IEEEtran}
\usepackage{xmpaper,lzqsymbol}
\usepackage{amssymb,amsmath,latexsym}
\usepackage{graphicx,subfigure}
\usepackage{color,cite,times}
\usepackage{epstopdf}
\usepackage{multirow}
\usepackage{array}

\usepackage{amsmath}
\usepackage{bm}
\usepackage[T1]{fontenc}

\usepackage{booktabs}
\usepackage{bbm} 
\usepackage{subfigure}
\usepackage{comment}
\usepackage{amsfonts}
\usepackage{subfigure}
\usepackage{booktabs}
\usepackage{algorithm}
\usepackage{algorithmic}
\usepackage{graphicx,dblfloatfix}
\usepackage{cite}
\usepackage{subeqnarray}
\usepackage{xcolor}
\usepackage{stmaryrd}
\usepackage{multirow}
\usepackage{diagbox}
\usepackage{url}
\usepackage{pifont}
\usepackage{hyperref}
\usepackage{upgreek}
\usepackage{booktabs}
\usepackage{lipsum}

\newcommand{\xiaowuhao}{\fontsize{9pt}{\baselineskip}\selectfont}



\begin{document}
\title{ Semi-Blind Channel-and-Signal Estimation for Uplink Massive MIMO With Channel Sparsity}
\author{Wenjing Yan and
	Xiaojun~Yuan,~\IEEEmembership{Senior~Member,~IEEE}\vspace{-0.3cm}
\thanks{W. Yan and X. Yuan  are with the Center for Intelligent Networking and Communications, the National Laboratory
of Science and Technology on Communications, the University of Electronic Science and Technology of China, Chengdu 611731,
China (e-mail: wjyan@std.uestc.edu.cn; xjyuan@uestc.edu.cn).
The work will be partially submitted for potential presentation in IEEE International Conference on Communications $\textrm{2019}$\cite{our}.}
}
\maketitle

\begin{abstract}
This paper considers the transceiver design for uplink massive multiple-input multiple-output (MIMO) systems with channel sparsity in the angular domain. Recent progress has shown that sparsity-learning-based blind signal detection is able to retrieve the channel and data by using massage-passing based sparse matrix factorization methods. Short pilots sequences are inserted into user packets to eliminate the so-called phase and permutation ambiguities inherent in sparse matrix factorization. In this paper, to exploit the knowledge of these short pilot sequences more efficiently, we propose a semi-blind channel-and-signal estimation (SCSE) scheme in which the knowledge of the pilot sequences are integrated into the message passing algorithm for sparse matrix factorization.
The SCSE algorithm involves enumeration over all possible user permutations, and so is time-consuming when the number of users is relatively large. To reduce complexity, we further develop the simplified SCSE (S-SCSE) to accommodate  systems with a large number of users.
 We show that our semi-blind signal detection scheme substantially outperforms the state-of-the-art blind detection and training-based schemes in the short-pilot regime.   
\end{abstract}

\section{Introduction}
Massive multiple-input multiple-output (MIMO) systems \cite{b1,b2,b3,b4} achieve significant performance improvement over the traditional communication systems in many aspects, such as increasing channel capacity, suppressing channel fading, and enhancing energy efficiency, etc. In a massive MIMO scenario, a base station (BS)
typically  equipped with an array of a few hundred antennas simultaneously serves many tens of terminals in a single time-frequency resource slot \cite{lu2014overview}. As the scale of the terminals or the array increases, the acquisition of channel state information (CSI) becomes one of the key obstacles for the utilization of massive MIMO \cite{HMT}.

Many studies have been attracted to the design of efficient and reliable techniques for channel acquisition. We are particularly focused on the uplink case, where users transmit signals to a BS [7]. In a training-based approach, each transmission frame is divided into two phases, namely, a training phase and a data transmission phase \cite{b5,b10}. In the training phase, pilots are transmitted to facilitate the estimation of the channel coefficients at the receiver side; in the data transmission phase, data are transmitted and the receiver performs detection based on the estimated channel.

Compared to separate signal processing for the two phases, joint channel-and-signal estimation is able to improve the system performance since partially detected data can be used as {\it soft} pilots to enhance the channel estimation accuracy in an iterative fashion \cite{ShiJin}. However, no matter whether separate or joint signal processing is employed, it is required in the training-based approach that the number of pilot symbols is no less than that of users, so as to ensure a vanishing channel estimation error \cite{b10,2003much}. As such, channel acquisition generally consumes a substantial portion of the system resource.

To reduce the channel acquisition overhead, another line of research is called the blind detection approach, in which the channel and data were estimated with little prior information of the signals from the transmitter side \cite{honig1995blind,wang1998blind,zheng2003}. In particular, it has been recently evidenced that a massive MIMO system exhibits channel sparsity in the angular domain, since signals usually impinge upon a massive antenna array from a limited range of angles \cite{S3,yin2013,muller2014,masood2015}. Based on the channel sparsity, a blind iterative detection technique \cite{b15} has been developed to avoid the use of pilots in channel acquisition. Approximate message passing algorithms \cite{Philip,b16} are used to simultaneously estimate the channel and data by factorizing the received noisy observation matrix.

Sparsity-learning based blind detection in \cite{b15}, however,  can be improved in a number of aspects. For example, the blind detection scheme in \cite{b15} imposes a relatively stringent requirement on the channel sparsity level and the signal-to-noise ratio (SNR) of the system to achieve a satisfactory error performance. More importantly, blind detection suffers from the so-called phase and permutation ambiguities inherent in matrix factorization. In \cite{b15}, a reference symbol and a user label are inserted in each user packet to eliminate the phase and permutation ambiguities after matrix factorization. Yet, as the reference symbols and the user labels (similar to pilots) are {\it a priori} known by the receiver, such knowledge can be integrated into the iterative process of matrix factorization to enhance the detection performance, rather than used for afterwards compensation.

To address the above issues, we propose a semi-blind detection scheme to jointly estimate the channel and the user signals in a sparse massive MIMO system. We focus on the scenario that a short pilot sequence is inserted into each user packet to assist the matrix factorization for joint channel and signal estimation. Here ``short pilot" means that the pilot sequence is not long enough to generate a relatively accurate initial channel estimate, and so the existing training-based approaches \cite{b5}\cite{ShiJin}  are unable to achieve a good performance. We show that, to efficiently exploit the short pilots, the phase and permutation ambiguities need to be skilfully estimated in the iterative process of the matrix factorization. As such, a message-passing based semi-blind channel and signal estimation (SCSE) algorithm is developed, building upon the framework of approximate message passing algorithms. The main contributions of this paper are summarized as follows.

\begin{itemize}
  \item {{\it SCSE algorithm for massive MIMO:}}
  We propose a novel semi-blind detection scheme for massive MIMO systems to jointly estimate the channel and the signals. The proposed semi-blind detection scheme is able to efficiently exploit the information of short pilots in the iterative process of sparse matrix factorization.
  \item {{\it Simplified SCSE algorithm for complexity reduction:}}
      The SCSE algorithm involves an exhaustive search of all possible user permutations, which is computationally infeasible when the number of users is relatively large. As such, we develop a simplified SCSE (S-SCSE) algorithm to avoid the burden of permutation enumeration.
  \item {}We show that our proposed S-SCSE is able to substantially outperform the state-of-the-art training-based and blind detection approaches \cite{ShiJin,b15} in the short-pilot regime. We also show that, compared to the SCSE algorithm, the S-SCSE algorithm significantly reduces the computational complexity while maintaining a similar performance, thereby striking a desirable balance between complexity and performance.
\end{itemize}

The remainder of this paper is organized as follows: Section~\ref{sec.sys} describes the sparse channel model and the system model for uplink MIMO systems. In Section~\ref{Semi-blind}, we formulate the joint channel and signal inference problem by including the estimation of the phase and permutation ambiguities inherent in sparse matrix factorization. Based on that, the SCSE and S-SCSE algorithms are derived based on the message-passing principles, and the selection metric for random initializations is described. Numerical results are presented in Section~\ref{sec.Num} to verify the effectiveness of our proposed algorithms. Finally, we conclude the paper in Section~\ref{sec.Con}.

{\it Notations:} Capital bold letters, lowercase bold letters, and regular letters represents matrices, vectors, and scalars, respectively.
For any matrix $\mathbf{A}$, $\mathbf{a}_i$ refers to the $i$th column of $\mathbf{A}$, and $a_{i,j}$ refers to the $(i,j)$th entry of $\mathbf{A}$. $\mathbb{C}$ denotes the complex field; $\mathcal{S}$ denotes a set; $\mathbf{P}$ denotes an arbitrary permutation matrix. For any set $\mathcal{S}$,  $|\mathcal{S}|$ represents the the cardinality of $\mathcal{S}$; $\mathbf{e}_i = [0,\ldots,0,1,0,\ldots,0]^{\rm T}$ with the only non-zero element being at the $i$th position; for any scalar $x$, $|x|$ represents the absolute value of $x$; $\|\cdot\|_2$ represents the $\ell_2$-norm.
The superscripts $(\cdot)^{\rm T}$, $(\cdot)^{\rm H}$, $(\cdot)^{-1}$ represent the  transpose, the conjugate transpose, and the inverse of a matrix, respectively; $\mathbb{E}(\cdot)$, $\delta(\cdot)$ and $e^{(\cdot)}$ represent the expectation, the Dirichlet function, and the exponential function; diag$\{\mathbf{a}\}$ represents the diagonal matrix with the diagonal specified by $\mathbf{a}$; $\lceil a \rceil$ represents the minimum integer larger than $a$. For any integer $\mathcal{I}_N$ denotes the set of integers from $1$ to $N$. $\mathcal{CN}(\cdot,\mu,\nu)$ represents a  complex circularly symmetric Gaussian distribution with the mean $\mu$ and covariance $\nu$.

\section{ System Model}
\label{sec.sys}
\subsection{Sparse Channel Model}

Consider an uplink massive MIMO system with $K$ single-antenna transmitters and a receiver equipped with $N$ antennas deployed as a uniform linear array (ULA), where $N\gg K\gg 1$. Denote by $\theta_{\ell,k}$ the angle of arrival (AoA) of the $\ell$th path from transmitter $k$.
The array steering vector for an incident signal from angle $\theta_{\ell,k}$ can be written as
\begin{align}
&\mathbf{a}_r(\theta_{\ell,k}) = \left[1,e^{-j2\pi \frac{ d}{\lambda}\cos(\theta_{\ell,k})}, \ldots, e^{- j2\pi \frac{(N-1)d}{\lambda}\cos(\theta_{\ell,k})}\right]^{\rm{T}}
\end{align}
where $d$ is the interval between any two adjacent receive antennas, and $\lambda$ is the wavelength of propagation.
We use the virtual representation method in \cite{S3} to divide the signal AoAs into $N$ resolution bins with the $k$th bin represented by $\theta_{\ell,k} = \arccos(\frac{\ell\lambda }{dN})$, $\ell \in \mathcal{I}_N \triangleq\{0,1,\ldots,N-1\}$.
Then, the physical channel of a massive MIMO system can be modeled as
\begin{align}
 \check{\mathbf{H}}=\mathbf{A}_r \mathbf{H}
\end{align}
where
$\mathbf{A}_r \!=\! \frac{1}{\sqrt{N}}\left[\mathbf{1},\mathbf{a}_r(\arccos\frac{\lambda}{dN}),\ldots,\mathbf{a}_r(\arccos\frac{(N-1)\lambda }{dN})\right]\!\!\!$ \\ $ \in \mathbb{C}^{N \times N}$
is the discrete Fourier transform (DFT) matrix, $\mathbf{1}$ is a $K$-dimensional all-one vector, and $\mathbf{H} = [\mathbf{h}_1, \mathbf{h}_2,\ldots,\mathbf{h}_K]$ is the projection of the physical channel in the angular domain.
Note that each $\mathbf{h}_{k}=[h_{1,k}, h_{2,k}, \ldots, h_{N,k}]^{\rm{T}}$ is the complex channel coefficient vector of user $k$, where $h_{n,k}$ is the aggregate channel coefficient in the resolution bin centered around $\theta_{\ell,k}$.

The physical channel of a massive MIMO system exhibits a sparse structure in the angular domain, since only a small portion of resolution bins receive electromagnetic waves from the transmitters. Therefore,
 the angular-domain channel representation $\mathbf{H}$ is a sparse matrix with a large portion of the elements being zero or close to zero
\cite{S3}.
Define the sparsity level of the massive MIMO channel by
\begin{align}
\rho = \frac{\vert \mathcal{S}\vert}{NK} < 1
\end{align}
where $\mathcal{S}$ is the support of $\mathbf{H}$,
and $\vert \mathcal{S}\vert$ represents the cardinality of $\mathcal{S}$.

We assume that each entry of $\mathbf{h}_{k}$ is independently drawn from a Bernoulli circularly symmetric complex Gaussian (B-CSCG) distribution
$(1-\rho)\delta(h)+\rho\mathcal{CN}(h;0,\sigma_{h,k}^2)$, where $\delta(\cdot)$ is the Dirac delta function, and
 $\sigma_{h,k}^2$ is the average power of the non-zero coefficients of the channel $\mathbf{h}_{k}$. Note that $\sigma_{h,k}^2$ is determined by the large-scale fading of user $k$, and is generally unknown to the receiver.\footnote{We assume that $\{\sigma_{h,k}^2\}$ are independently drawn from a certain known distribution. The distribution will be specified in the simulations in Section~\ref{sec.Num}. The estimation of $\{\sigma_{h,k}^2\}$ will be discussed later in Section~\ref{Para-tun}.}
Based on the above discussion, the distribution of the channel is given by
\begin{align} \label{channel.H}
 P_{\mathbf{H}}(\mathbf{H})=\prod_{n=1}^N\prod_{k=1}^K (1-\rho)\delta(h_{n,k})+\rho\mathcal{CN}(h_{n,k};0,\sigma_{h,k}^2).
\end{align}

\subsection{System Model}

Assume that the channel is block-fading with coherence time $T$. The massive MIMO system in the angular domain over $T$ time slots can be modeled as
\begin{align} \label{System}
\mathbf{Y} = \mathbf{H}\mathbf{X} + \mathbf{W} = \mathbf{Z}+ \mathbf{W}
\end{align}
where $\mathbf{Y} \in \mathbb{C}^{N\times T}$ is the transformed observation matrix in the angular domain,
$\mathbf{W} \in \mathbb{C}^{N\times T}$ is an additive white Gaussian noise (AWGN) with each entry independently drawn from $\mathcal{CN}(0,N_0)$ with $N_0$ being the noisy power, $\mathbf{H}\in \mathbb{C}^{N\times K}$ is the angular-domain channel matrix as aforementioned,
$\mathbf{X}=[\mathbf{x}_1, \mathbf{x}_2,\ldots, \mathbf{x}_K]^{\rm T}\in \mathbb{C}^{K\times T}$ is the signal matrix, and $\mathbf{Z}=\mathbf{H}\mathbf{X}\in \mathbb{C}^{N\times T}$.
Each entry of $\mathbf{X}$ is modulated by using a constellation $\mathcal{C}=\{c_1,c_2,\ldots,c_{\vert \mathcal{C}\vert}\}$, where $\vert \mathcal{C}\vert$ is the cardinality of $\mathcal{C}$. That is, $x_{k,t}$ is uniformly drawn from $\mathcal{C} $ for $\forall k,t$, where $x_{k,t}$ is the $t$th entry of $\mathbf{x}_k$.
Assume that $\mathcal{C}$ is rotationally invariant for any rotation angle $\theta \in \Omega$, where $\Omega=\{\omega_1,\omega_2,\ldots,\omega_{\vert \Omega\vert}\}$ is an angle set. For example, $\Omega=\{0^\circ, 90^\circ,180^\circ,270^\circ\}$ for standard quadrature amplitude modulation (QAM).
For each user $k$, the first $T_{\rm {P}}$ symbols of $\mathbf{x}_k$ (denoted by ${\mathbf{x}_{{\rm P},k}}\in \mathbb{C}^{P \times 1}$) are assigned as pilots, and the remaining $T-T_{\rm P}$ are data symbols.
 We use $\mathcal{T}_P$ to represent the set $\{1,2,\ldots, T_{\rm P}\}$, and $\mathcal{T}_D$ to represent the set $\{T_{\rm P}+1,T_{\rm P}+2,\ldots, T\}$.
 Let $\mathbf{X}_{\rm P}=[\mathbf{x}_{{\rm P},1}, \mathbf{x}_{{\rm P},2}, \ldots ,\mathbf{x}_{{\rm P},K}]^{\rm{T}}$ be the pilot matrix occupying the first $T_{\rm {P}}$ columns of $\mathbf{X}$, and
let $\mathbf{X}_{\rm D}=[\mathbf{x}_{{\rm D},1}, \mathbf{x}_{{\rm {D}},2}, \ldots ,\mathbf{x}_{{\rm D},K}]^{\rm{T}}$ be the data matrix occupying the remaining $T-T_{\rm {P}}$ columns of $\mathbf{X}$, i.e., $\mathbf{X}=[\mathbf{X}_{\rm {P}},\mathbf{X}_{\rm {D}}]$.
Similarly, $\mathbf{Y}$ can be expressed as $\mathbf{Y}=[\mathbf{Y}_{\rm {P}},\mathbf{Y}_{\rm {D}}]$, where $\mathbf{Y}_{\rm {P}}$ and $\mathbf{Y}_{\rm {D}}$ correspond to $\mathbf{X}_{\rm {P}}$ and $\mathbf{X}_{\rm {D}}$, respectively.
Assume that the entries of $\mathbf{X}$ are independent of each other, i.e.,
\begin{align} \label{data.X}
P_{\mathbf{X}}(\mathbf{X})
= \prod_{k=1}^K\prod_{t=1}^T p_{x_{k,t}}(x_{k,t}).
\end{align}
Denote by $P$ the total power budget of the transmitters and $\alpha_k P$ the average transmission power of the $k$th transmitter. Then, each transmitter is power-constrained as
\begin{align}
\frac{1}{T}\mathbb{E}[\mathbf{x}_k^{\rm{H}} \mathbf{x}_k]\leqslant \alpha_k P, \quad \textrm{for all}~ k\in {\mathcal{I}}_K\triangleq\{1,2,\ldots,K\}
\end{align}
where $\alpha_k\geqslant 0$ for $k\in {\mathcal{I}}_K$ with $\sum_{k=1}^{K}\alpha_k=1$.

\section{Problem Description}
\label{Semi-blind}

In this paper, our goal is to retrieve both the channel matrix $\mathbf{H}$ and the symbol matrix $\mathbf{X}_{\rm D}$ from the observed data matrix $\mathbf{Y}$ together with the prior knowledge of the pilot matrix $\mathbf{X}_{\rm P}$.
 This problem can be formulated by using the {\it maximum a posteriori} (MAP) principle as
\begin{align}\label{joint.for}
\left( \hat{\mathbf{H}},\hat{\mathbf{X}}_{\rm D}\right) = \arg\max_{\mathbf{H},\mathbf{X}_{\rm D}}p_{\mathbf{H},\mathbf{X}_{\rm D}|\mathbf{Y}, \mathbf{X}_{\rm P}}(\mathbf{H},\mathbf{X}_{\rm D}|\mathbf{Y},\mathbf{X}_{\rm P})
\end{align}
where $\hat{\mathbf{H}}$ and $\hat{\mathbf{X}}_{\rm D}$ are the estimates of the channel matrix $\mathbf{H}$ and the signal matrix
$\mathbf{X}_{\rm D}$, respectively.

A straightforward approach to solving \eqref{joint.for} is first to estimate the channel $\mathbf{H}$ based on $\mathbf{Y}_{\rm{P}}$ (with known $\mathbf{X}_{\rm P}$) and then to estimate the data matrix
$\mathbf{X}_{\rm{D}}$ based on $\mathbf{Y}_{\rm{D}}$ and the channel estimate $\hat{\mathbf{H}}$. This approach is referred to as the separate channel-and-signal estimation method scheme in the following. In principle, the estimated data can be used to further refine the channel estimate and hence improve the system performance. To this end, the authors in \cite{ShiJin} proposed a joint channel and signal estimation method which involves approximate message passing over the factor graph obtained by factorizing the probability distribution $p_{\mathbf{H},\mathbf{X}_{\rm D}|\mathbf{Y}, \mathbf{X}_{\rm P}}$ in \eqref{joint.for}.

In this paper, we focus on the ``semi-blind'' scenario where the pilot length $T_{\rm P}$ is not large enough to provide a relatively accurate initial channel estimate for data detection. In this scenario, both the separate and joint training-based estimation approaches discussed above do not work well.
 The main reason is that when $T_{\rm P}$ is small, the scheme in \cite{ShiJin} is close to blind channel-and-signal estimation in which the issue of phase and permutation ambiguities arise \cite{b15}. For a small $T_{\rm P}$, the knowledge of $\mathbf{X}_{\rm{P}}$ is not ``strong'' enough to correct the phase and permutation ambiguities in the iterative estimation process. As a result,
 the scheme in \cite{ShiJin} may perform even worse than the blind detection method in \cite{b15}.

Instead, we aim to find a method that can efficiently exploit the knowledge of $\mathbf{X}_{\rm{P}}$ in the presence of phase and permutation ambiguities. To this end, we need to estimate the phase and permutation ambiguities in the iterative process of message passing. Denote by $\mathbf{\Pi}=[\boldsymbol{\pi}_1,\boldsymbol{\pi}_2,\ldots,\boldsymbol{\pi}_K]^{\rm T}\in \mathbb{C}^{K\times K}$ an arbitrary permutation matrix,
and by $\mathbf{\Sigma}=\diag\{\sigma_1,\sigma_2,\ldots,\sigma_K\}$
 a diagonal matrix where each phase shift $\sigma_i = e^{j\omega_i}$ with $\omega_i$ independently and uniformly taken from $\Omega$.
It is known in \cite{b15} that blind detection suffers from the phase and permutation ambiguities, i.e.,
if  $(\hat{\mathbf{H}}, \hat{\mathbf{X}})$ is a valid factorization given $\mathbf{Y}$ in \eqref{System}, then $(\hat{\mathbf{H}}, {\mathbf{\Pi}}^{-1}{\mathbf{\Sigma}}^{-1}, {\mathbf{\Sigma}}{\mathbf{\Pi}}\hat{\mathbf{X}})$ is also a valid factorization.
We will estimate $\mathbf{\Pi}$ and $ \mathbf{\Sigma}$ in the message passing process by exploiting the knowledge of $\mathbf{X}_{\rm P}$.
To this end, we define auxiliary variables
 \begin{align} \label{H}
 \tilde{\mathbf{H}}=\mathbf{H}{\mathbf{\Pi}}^{-1}{\mathbf{\Sigma}}^{-1} \quad \text{and}   \quad    \tilde{\mathbf{X}}={\mathbf{\Sigma}}{\mathbf{\Pi}}\mathbf{X}.
 \end{align}
 
 Recall from \eqref{channel.H} that the columns of $\mathbf{H}$ are independently and identically distributed (i.i.d.). Thus, for any permutation $\mathbf{\Pi}$, $\mathbf{H}\mathbf{\Pi}^{-1}$ has the same distribution as $\mathbf{H}$ does. Then,
 \begin{align}
 p(\mathbf{H}\mathbf{\Pi}^{-1},\mathbf{\Pi})=p(\mathbf{H}\mathbf{\Pi}^{-1}|\mathbf{\Pi})p(\mathbf{\Pi})
 =p(\mathbf{H}\mathbf{\Pi}^{-1})p(\mathbf{\Pi}).
 \end{align}
 That is, $\mathbf{H}\mathbf{\Pi}^{-1}$ is independently of $\mathbf{\Pi}$. Similarly, for any phase ambiguity matrix $\mathbf{\Sigma}$,
 $\tilde{\mathbf{H}}=\mathbf{H}{\mathbf{\Pi}}^{-1}{\mathbf{\Sigma}}^{-1}$ has the same distribution as $\mathbf{H}$ does. Therefore,
 $\tilde{\mathbf{X}}$ is independent of $\mathbf{\Pi}$ and $\mathbf{\Sigma}$. Since $\mathbf{H}$, $\mathbf{X}$, $\mathbf{\Pi}$, and $\mathbf{\Sigma}$ are independent of each other, we conclude that $\tilde{\mathbf{H}}$, $\tilde{\mathbf{X}}$, $\mathbf{\Pi}$, and $\mathbf{\Sigma}$ are also independent of each other, i.e.,
\begin{align} \label{HXCP}
&p_{\tilde{\mathbf{H}},\tilde{\mathbf{X}},\mathbf{\Pi},\mathbf{\Sigma}}(\tilde{\mathbf{H}},\tilde{\mathbf{X}},\mathbf{\Pi},\mathbf{\Sigma})
= p_{\tilde{\mathbf{H}}}(\tilde{\mathbf{H}}) p_{\tilde{\mathbf{X}}}(\tilde{\mathbf{X}}) p_{\mathbf{\Pi}}(\mathbf{\Pi}) p_{\mathbf{\Sigma}}(\mathbf{\Sigma}).
\end{align}
Then we recast the problem in \eqref{joint.for} as
\begin{align} \label{Sys2}
 \max_{\tilde{\mathbf{H}},\tilde{\mathbf{X}}, \mathbf{\Sigma},\mathbf{\Pi}} p_{\tilde{\mathbf{H}},\tilde{\mathbf{X}},\mathbf{\Sigma},\mathbf{\Pi}|\mathbf{Y},\mathbf{X}_{\rm P}}
(\tilde{\mathbf{H}},\tilde{\mathbf{X}},\mathbf{\Sigma},\mathbf{\Pi}|\mathbf{Y},\mathbf{X}_{\rm P}).
\end{align}
 From the Bayes' rule, we obtain
 \begin{figure*}
  \centering
  \includegraphics[width=6 in]{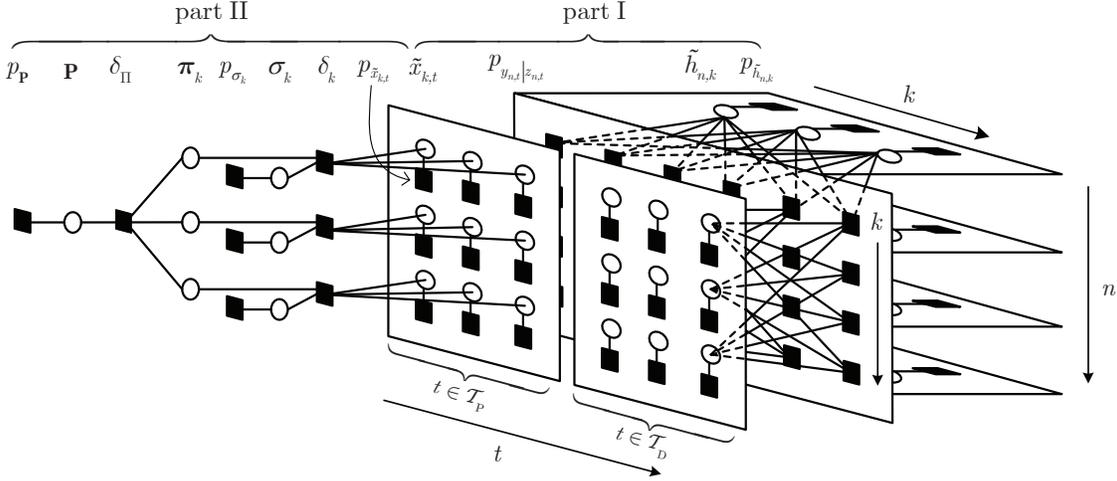}\\
  \caption{The factor graph representation for the joint
  probability in \eqref{joint.pro2} with  $N=4$, $K=3$, $T_{\rm P}=3$, and $T=6$.}\label{figure_factor}
\end{figure*}
\begin{subequations} \label{joint.pro}
\begin{align}
&~~~~p_{\tilde{\mathbf{H}},\tilde{\mathbf{X}},\mathbf{\Sigma},\mathbf{\Pi}|\mathbf{Y},\mathbf{X}_P}(\tilde{\mathbf{H}},
\tilde{\mathbf{X}},\mathbf{\Sigma},\mathbf{\Pi}|\mathbf{Y},\mathbf{X}_P)\notag\\
&=\frac{p_{\mathbf{Y},\mathbf{X}_{\rm P},\tilde{\mathbf{H}},\tilde{\mathbf{X}},\mathbf{\Sigma},\mathbf{\Pi}}
(\mathbf{Y},\mathbf{X}_{\rm P},\tilde{\mathbf{H}},\tilde{\mathbf{X}},\mathbf{\Sigma},\mathbf{\Pi})}
{p_{\mathbf{Y},\mathbf{X}_{\rm P}}(\mathbf{Y},\mathbf{X}_{\rm P})} \label{joint.pro.a}\\
&\propto p_{\mathbf{Y}|\mathbf{X}_{\rm P},\tilde{\mathbf{H}},\tilde{\mathbf{X}},\mathbf{\Sigma},\mathbf{\Pi}} (\mathbf{Y}|\mathbf{X}_{\rm P},\tilde{\mathbf{H}},\tilde{\mathbf{X}},\mathbf{\Sigma},\mathbf{\Pi})\notag\\
&~~~\times p_{\mathbf{X}_{\rm P}|\tilde{\mathbf{H}},\tilde{\mathbf{X}},\mathbf{\Sigma},\mathbf{\Pi}}
(\mathbf{X}_{\rm P}|\tilde{\mathbf{H}},\tilde{\mathbf{X}},\mathbf{\Sigma},\mathbf{\Pi})   p_{\tilde{\mathbf{H}},\tilde{\mathbf{X}},\mathbf{\Sigma},\mathbf{\Pi}}
({\tilde{\mathbf{H}}},\tilde{\mathbf{X}},\mathbf{\Sigma},\mathbf{\Pi})\label{joint.pro.b}
\end{align}
\end{subequations}
\begin{subequations} \label{joint.pro}
\begin{align}
& = p_{\mathbf{Y}|\tilde{\mathbf{H}},\tilde{\mathbf{X}}}(\mathbf{Y}|\tilde{\mathbf{H}},\tilde{\mathbf{X}})
p_{\mathbf{X}_{\rm P}|\tilde{\mathbf{X}},\mathbf{\Sigma},\mathbf{\Pi}}
(\mathbf{X}_{\rm P}|\tilde{\mathbf{X}},\mathbf{\Sigma},\mathbf{\Pi}) \notag\\
&~~~\times p_{\tilde{\mathbf{H}}}({\tilde{\mathbf{H}}}) p_{\tilde{\mathbf{X}}}({\tilde{\mathbf{X}}}) p_{\mathbf{\Sigma}}(\mathbf{\Sigma})p_{\mathbf{\Pi}}(\mathbf{\Pi}) ~~~~~~~~~~~~~~~~~~~~~~~~~~~\textrm{(12c)}  \notag\\
& = \left[\prod_n\prod_t p_{y_{n,t}|z_{n,t}}\left(y_{n,t}|z_{n,t}=\sum_k \tilde{h}_{n,k}\tilde{x}_{k,t}\right)\right]\notag\\
&~~~\times \left[\prod_n\prod_k p_{\tilde{h}_{n,k}}(\tilde{h}_{n,k})\right] \left[\prod_k\prod_t p_{\tilde{x}_{k,t}}(\tilde{x}_{k,t})\right]\notag\\
&~~~\times \delta(\tilde{\mathbf{X}}_{{\rm P}}-\mathbf{\Sigma} \mathbf{\Pi} {\mathbf{X}}_{\rm P})
\left[\prod_k p_{\sigma_{k}}(\sigma_{k})\right] p_{\mathbf{\Pi}}(\mathbf{\Pi})~~~~~~~~~~~~~\textrm{(12d)}  \notag
\end{align}
\end{subequations}
where \eqref{joint.pro.a} follows from the Bayes' rule; the notation $\propto $ in step \eqref{joint.pro.b} means equality up to a constant scaling factor; $\textrm{(12c)}$ is from the facts that (\expandafter{\romannumeral1}) $(\tilde{\mathbf{X}}_{\rm P},\mathbf{\Sigma},\mathbf{\Pi})\rightarrow (\tilde{\mathbf{H}},\tilde{\mathbf{X}}) \rightarrow \mathbf{Y}$ forms a Markov chain, (\expandafter{\romannumeral2}) $\mathbf{X}_{\rm{P}}$ is independent of $\tilde{\mathbf{H}}$ for any given $\tilde{\mathbf{X}}$, $\mathbf{\Sigma}$, and $\mathbf{\Pi}$, and (\expandafter{\romannumeral3}) $\tilde{\mathbf{H}}$, $\tilde{\mathbf{X}}$, $\mathbf{\Sigma}$, $\mathbf{\Pi}$ are independent of each other by \eqref{HXCP};  $p_{\sigma_k}(\sigma_k)$ in $\textrm{(12d)}$ is the probability density of the phase shift of user $k$.
 Recall that $p_{\sigma_k}(\sigma_k)=\frac{1}{|\Omega|}\sum_{\omega \in \Omega}\delta(\sigma_k-e^{j\omega})$, where $|\Omega|$ is the cardinality of $\Omega$.
The factorization in $\textrm{(12)}$ will be used in the development of message passing algorithms in the subsequent section.
We will show that the message passing algorithms developed based on $\textrm{(12)}$ performs much better than the one based on the factorization of
$p_{\mathbf{H},\mathbf{X}_{\rm D}|\mathbf{Y}, \mathbf{X}_{\rm P}}$ as in \cite{ShiJin}

\section{Semi-Blind Message Passing Algorithms }

\subsection{Factor Graph Representation}
\label{Sec.S-SBDA}

To describe the massage passing process more clearly, we introduce an auxiliary variable $\mathbf{P}\in \mathbb{C}^{K\times K}$ to denote a random permutation matrix, where $\mathbf{P}\in \mathcal{P} \triangleq \{\mathbf{P}_1,\mathbf{P}_2,\cdots,\mathbf{P}_{K!}\}$ with an equal probability with $\mathcal{P}$ being the set of all permutations.\footnote{It is possible to design message passing directly based on the factorization in (12). However, the introduction of $\mathbf{P}$ yields a unified view of the derivations of the SCSE and S-SCSE algorithms. In particular, we show in Section ~\ref{Sec.S-SCSE} that the S-SCSE algorithm is derived simply by deleting the constraint $\delta_{\mathbf{\Pi}}$ in Fig.~\ref{figure_factor}.}
Then, the fact that $\mathbf{\Pi}$ is a random permutation can be represented by the following joint distribution:
\begin{align}
p_{\mathbf{\Pi},\mathbf{P}}(\mathbf{\Pi},\mathbf{P})= p_{\mathbf{P}}(\mathbf{P})\delta\left(\left[\boldsymbol{\pi}_1,\boldsymbol{\pi}_2,\ldots,\boldsymbol{\pi}_K \right]^{\rm T} - \mathbf{P}\right)  \label{pi.1}
\end{align}
where $p_{\mathbf{P}}(\mathbf{P})=\frac{1}{k!}\sum_{\ell=1}^{K!}\delta(\mathbf{P}-\mathbf{P}_\ell)$, $\boldsymbol{\pi}_k^{\rm T}$ is taken from the set $\{ \mathbf{e}_i\}_{i=1}^K$ with $\mathbf{e}_i = [0,\ldots,0,1,0,\ldots,0]^{\rm T}$ being the $i$th column of the $K$-by-$K$ identity matrix.
With the inclusion of the auxiliary variable $\mathbf{P}$, the factorization in \eqref{joint.pro} converts to
\begin{align}
&~~~~p_{\tilde{\mathbf{H}},\tilde{\mathbf{X}},\mathbf{\Sigma},\mathbf{\Pi},\mathbf{P}|\mathbf{Y},\mathbf{X}_P}
(\tilde{\mathbf{H}},\tilde{\mathbf{X}},\mathbf{\Sigma},\mathbf{\Pi},\mathbf{P}|\mathbf{Y},\mathbf{X}_P)\notag\\
& = \left[\prod_n\prod_t p_{y_{n,t}|z_{n,t}}\left(y_{n,t}|z_{n,t}=\sum_k \tilde{h}_{n,k}\tilde{x}_{k,t}\right)\right]\notag\\
&~~~\times \left[\prod_n\prod_k p_{\tilde{h}_{n,k}}(\tilde{h}_{n,k})\right] \left[\prod_k\prod_t p_{\tilde{x}_{k,t}}(\tilde{x}_{k,t})\right]\notag\\
&~~~\times \left[\prod_{k=1}^{K} \delta(\tilde{\mathbf{x}}_{{\rm P},k}-\sigma_k \boldsymbol{\pi}_k^{\rm T} {\mathbf{X}}_{\rm P})\right]
\left[\prod_k p_{\sigma_{k}}(\sigma_{k})\right] \notag\\
&~~~\times p_{\mathbf{P}}(\mathbf{P})\delta\left(\left[\boldsymbol{\pi}_1,\boldsymbol{\pi}_2,\ldots,\boldsymbol{\pi}_K \right]^{\rm T} - \mathbf{P}\right).
\label{joint.pro2}
\end{align}
 The factorized posterior distribution in \eqref{joint.pro2} can be represented by a factor graph, as depicted in Fig.~\ref{figure_factor}.
 In Fig.~\ref{figure_factor}, we use a brief form of $\delta_{k}$ to represent
 $\delta(\tilde{\mathbf{x}}_{{\rm P}, k}-\sigma_k\boldsymbol{\pi}_k^{\rm T} {\mathbf{X}}_{\rm P})$, $k\in \mathcal{I}_K$,  and $\delta_{\mathbf{\Pi}}$ to represent $\delta([\boldsymbol{\pi}_1,\boldsymbol{\pi},\ldots,\boldsymbol{\pi}_K]^{\rm T}-\mathbf{P})$.
Each hollow circle in Fig.~\ref{figure_factor} represents a ``variable node'' corresponding to a random variable involved in \eqref{joint.pro2}, and each black solid square represents a ``factor node'' corresponding to a factor function in \eqref{joint.pro2}.
A variable node is connected to a factor node if the  variable appears in the  factor function.

In Fig.~\ref{figure_factor}, we divide the whole factor graph into two parts.
In part \uppercase\expandafter{\romannumeral1}, we estimate the channel matrix $\tilde{\mathbf{H}}$ and signal matrix $\tilde{\mathbf{X}}$ based on $\mathbf{Y}$ and the knowledge that $\tilde{\mathbf{H}}$ is sparse;
in part \uppercase\expandafter{\romannumeral2}, we use the knowledge of $\mathbf{X}_{\rm{P}}$ to improve the estimation of $\tilde{\mathbf{X}}_{\rm{P}}$ and estimate $\mathbf{\Pi}$ and $\mathbf{\Sigma}$ based on the constraint of $\tilde{\mathbf{X}}_{{\rm P}}=\mathbf{\Sigma} \mathbf{\Pi} {\mathbf{X}}_{\rm P}$.
We derive the semi-blind detection algorithm based on the message passing principles over the factor graph in Fig.~\ref{figure_factor}.
Note that the constraints in part \uppercase\expandafter{\romannumeral1} are related to factorizing the matrix product $\tilde{\mathbf{H}}\tilde{\mathbf{X}}$. This part can be realized by following the BiG-AMP algorithm in \cite{b16}. Therefore, in what follows, we focus on the derivation of the message passing algorithm for part \uppercase\expandafter{\romannumeral2}.

\subsection{SCSE Algorithm}

The messages involved in part \uppercase\expandafter{\romannumeral2} are described as follows.
Denote by $\Delta_{a\rightarrow b}(\cdot)$ the message from node $a$ to node $b$ and by $\Delta_{c}(\cdot)$ the marginal posterior of variable $c$.

$1)$ The message from $\tilde{x}_{k,t}$ to $\delta_k$ is given by
\begin{align}
\Delta_{\tilde{x}_{k,t} \rightarrow \delta_k} (\tilde{x}_{k,t})\varpropto  p_{\tilde{x}_{k,t}}(\tilde{x}_{k,t})
\prod_{n=1}^N \Delta_{p_{y_{n,t}|z_{n,t}}\rightarrow \tilde{x}_{k,t}}(\tilde{x}_{k,t}).\label{MP.1}\!\!\!\!
\end{align}

$2)$ The message from $\sigma_k$ to $\delta_k$ is given by
\begin{align}
\Delta_{\sigma_k \rightarrow \delta_k} (\sigma_k)= p_{\sigma_k}(\sigma_k)=\frac{1}{|\Omega|}\sum_{\omega \in \Omega}\delta(\sigma_k-e^{j\omega})\label{MP.2}.
\end{align}

$3)$ The message from $\delta_k$ to $\boldsymbol{\pi}_k$ is given by
\begin{align}
&\Delta_{\delta_k \rightarrow \boldsymbol{\pi}_k} (\boldsymbol{\pi}_k)\notag\\
&\varpropto  \int_{\sigma_k,\{x_{k,t}\}_{t=1}^{T_{\rm P}} }
\delta(\tilde{\mathbf{x}}_{{\rm P}, k}-{\sigma}_k \boldsymbol{\pi}_k^{\rm T} {\mathbf{X}}_{\rm P})\notag\\
&~~~~~~~~~~\times p_{\sigma_k}(\sigma_k) \prod_{t=1}^{T_{\rm P}}  \Delta_{\tilde{x}_{k,t} \rightarrow \delta_k}(\tilde{x}_{k,t}). \label{MP.3}
\end{align}
Since $\tilde{x}_{k,t}$, $\boldsymbol{\pi}_k$, and $\sigma_k$ are discrete variables, we write the message in \eqref{MP.3} in its discrete form as
\begin{align}
&P_{\delta_k \rightarrow \boldsymbol{\pi}_k} (\boldsymbol{\pi}_k=\mathbf{e}_i)\notag\\
&= \frac{1}{C} \sum_{\omega \in \Omega}\prod_{t=1}^{T_{\rm P}} P_{ \tilde{x}_{k,t}\rightarrow \delta_k }(\tilde{x}_{k,t}= e^{j\omega} x_{i,t}),
\quad i \in \mathcal{I}_K \label{MP.4}
\end{align}
where $P_{\tilde{x}_{k,t}\rightarrow \delta_k}(\tilde{x}_{k,t}= e^{j\omega} x_{i,t})$ denotes the probability of $\tilde{x}_{k,t}= e^{j\omega} x_{i,t}$ specified in the message $\Delta_{\tilde{x}_{k,t}\rightarrow \delta_k}(\tilde{x}_{k,t})$, and $C$ is a generic normalization factor.
Clearly, the message from $\boldsymbol{\pi}_k$ to $\delta_{\mathbf{\Pi}}$ is
\begin{align}
\Delta_{ \boldsymbol{\pi}_k \rightarrow \delta_{\mathbf{\Pi}}} (\boldsymbol{\pi}_k)= \Delta_{\delta_k \rightarrow \boldsymbol{\pi}_k}(\boldsymbol{\pi}_k).  \label{MP.5}
\end{align}

$4)$ The message from $\mathbf{P}$ to $\delta_{\mathbf{\Pi}}$ is given by
\begin{align}
\Delta_{\mathbf{P} \rightarrow \delta_{\mathbf{\Pi}}} (\mathbf{P})= p_{\mathbf{P}}(\mathbf{P})=\frac{1}{K!}\sum_{\ell=1}^{K!}\delta(\mathbf{P}-\mathbf{P}_\ell).
\end{align}

$5)$ Combining the message from $\mathbf{P}$ to $\delta_{\mathbf{\Pi}}$ and the messages from $\{\boldsymbol{\pi}_k'\}_{k'= \neq k}^K$ to $\delta_{\mathbf{\Pi}}$, we obtain
\begin{align}
&\Delta_{\delta_{\mathbf{\Pi}} \rightarrow \boldsymbol{\pi}_k} (\boldsymbol{\pi}_k)\notag\\
&\varpropto  \int_{\mathbf{P}, \{\boldsymbol{\pi}_{k'}\}_{k'=1 \neq k}^K} \delta([\boldsymbol{\pi}_1,\boldsymbol{\pi}_2,\ldots,\boldsymbol{\pi}_K]^{\rm T}-\mathbf{P}) \notag\\
&~~~~~~~~~~~~~~\times \prod_{k'=1 \neq k}^K \Delta_{ \boldsymbol{\pi}_{k'} \rightarrow \delta_{\mathbf{\Pi}}} (\boldsymbol{\pi}_{k'})p_{\mathbf{P}}(\mathbf{P}). \label{MP.6}
\end{align}
Denote by $\mathbf{p}_{\ell,k}$ the transpose of the $k$th row of $\mathbf{P}_{\ell}$. The discrete form of the above message can be written as
\begin{align}
&P_{\delta_{\mathbf{\Pi}} \rightarrow \boldsymbol{\pi}_k} (\boldsymbol{\pi}_k=\mathbf{e}_i) \notag\\
&= \frac{1}{C} \sum_{\substack{\ell=1\\ \ell:~ \mathbf{p}_{\ell,k}=\mathbf{e}_i }}^{K!} \prod_{k'=1 \neq k}^K P_{ \boldsymbol{\pi}_{k'} \rightarrow \delta_{\mathbf{\Pi}}}(\boldsymbol{\pi}_{k'}=\mathbf{p}_{\ell,k'}),  \notag\\
&~~~~~~~~~~~~~~~~~~~~~~~~~~~~~~~~~~~~~~~~~~~~~~i \in \mathcal{I}_K   \label{MP.7}
\end{align}
where $P_{ \boldsymbol{\pi}_{k'} \rightarrow \delta_{\mathbf{\Pi}}} (\boldsymbol{\pi}_{k'}=\mathbf{p}_{\ell,k'})$ denotes the probability of $\boldsymbol{\pi}_{k'}=\mathbf{p}_{\ell,k'}$ specified by the message $\Delta_{\boldsymbol{\pi}_{k'}\rightarrow \delta_{\mathbf{\Pi}}}(\boldsymbol{\pi}_{k'})$.
Similar to \eqref{MP.5}, the message from $\boldsymbol{\pi}_k$ to $\delta_k$ is
\begin{align}
\Delta_{ \boldsymbol{\pi}_k \rightarrow \delta_k} (\boldsymbol{\pi}_k)= \Delta_{\delta_{\mathbf{\Pi}} \rightarrow \boldsymbol{\pi}_k}(\boldsymbol{\pi}_k).  \label{MP.8}
\end{align}

$6)$ The message from $\delta_k$ to $\tilde{x}_{k,t}$ is given by
\begin{align}
&\Delta_{\delta_k \rightarrow \tilde{x}_{k,t}}(\tilde{x}_{k,t}) \notag\\
&\varpropto  \int_{\boldsymbol{\pi}_k,\sigma_k, \{x_{k,t'}\}_{t'=1\neq t}^{T_{\rm P}} } \delta(\tilde{\mathbf{x}}_{{\rm P}, k}-\sigma_k \boldsymbol{\pi}_k^{\rm T} {\mathbf{X}}_{\rm P}) \notag\\
&~~~\times p_{\sigma_k}(\sigma_k) \Delta_{\boldsymbol{\pi}_k \rightarrow \delta_k} (\boldsymbol{\pi}_k) \prod_{t'=1 \neq t}^{T_{\rm P}} \Delta_{ \tilde{x}_{k,t'}\rightarrow \delta_k }(\tilde{x}_{k,t'}).  \label{MP.9}
\end{align}
We write the above message in its discrete form as
\begin{align}
&P_{\delta_k \rightarrow \tilde{x}_{k,t}}(\tilde{x}_{k,t}=c) \notag\\
&= \frac{1}{C}\sum_{\substack{i\in \mathcal{I}_K, \omega\in \Omega\\ \text{for~} e^{j\omega}x_{i,t}=c}}
P_{\delta_{\mathbf{\Pi}} \rightarrow \boldsymbol{\pi}_k} (\boldsymbol{\pi}_k=\mathbf{e}_i)\notag\\
&~~~\times \prod_{t'=1 \neq t}^{T_{\rm P}} P_{ \tilde{x}_{k,t'}\rightarrow \delta_k }(\tilde{x}_{k,t'}=e^{j\omega}x_{i,t'}),\quad c \in \mathcal{C}.
\end{align}

$7)$ The message from $\tilde{x}_{k,t}$ to $p_{y_{n,t}|z_{n,t}}$ is given by
\begin{align}
&\Delta_{\tilde{x}_{k,t} \rightarrow p_{y_{n,t}|z_{n,t}}}(\tilde{x}_{k,t}) \notag\\
&\varpropto  p_{\tilde{x}_{k,t}}(\tilde{x}_{k,t}) \Delta_{\delta_k \rightarrow \tilde{x}_{k,t}}(\tilde{x}_{k,t}) \notag\\
&~~~ \times \prod_{n'=1\neq n}^N \Delta_{p_{y_{n',t}|z_{n',t}} \rightarrow \tilde{x}_{k,t}}(\tilde{x}_{k,t}). \label{MP.xp}
\end{align}
where $p_{\tilde{x}_{k,t}}(\tilde{x}_{k,t}) = p_{x_{k,t}}(\tilde{x}_{k,t})$ is the prior distribution of $\tilde{x}_{k,t}$ determined by the modulation of the data, and $\Delta_{\delta_k \rightarrow \tilde{x}_{k,t}}(\tilde{x}_{k,t})$ is the message provided by the prior knowledge of $\mathbf{X}_{\rm P}$.
The massage passing process of part II can be realized by the BiG-AMP algorithm in \cite{b16}.
Note that the marginal posterior $\Delta_{\tilde{x}_{k,t}}(\tilde{x}_{k,t})= \Delta_{\delta_k \rightarrow \tilde{x}_{k,t}}(\tilde{x}_{k,t})
\Delta_{\tilde{x}_{k,t} \rightarrow \delta_k}(\tilde{x}_{k,t})$ rather then $\Delta_{\tilde{x}_{k,t} \rightarrow p_{y_{n,t}|z_{n,t}}}(\tilde{x}_{k,t})$ is needed in the BiG-AMP algorithm for complexity reduction.
The discrete form of $\Delta_{\tilde{x}_{k,t}}(\tilde{x}_{k,t})$ is given by
\begin{align}
&P_{\tilde{x}_{k,t}} (\tilde{x}_{k,t}=c)\notag\\
&\varpropto P_{\delta_k \rightarrow \tilde{x}_{k,t}}(\tilde{x}_{k,t}=c)
P_{\tilde{x}_{k,t} \rightarrow \delta_k}(\tilde{x}_{k,t}=c) \notag\\
&= \frac{1}{C}\sum_{\substack{i\in \mathcal{I}_K, \omega\in \Omega\\ \text{for:~} e^{j\omega}x_{i,t}=c}}
P_{\delta_{\mathbf{\Pi}} \rightarrow \boldsymbol{\pi}_k} (\boldsymbol{\pi}_k=\mathbf{e}_i)\notag\\
&~~~\times \prod_{t'=1 \neq t}^{T_{\rm P}} P_{ \tilde{x}_{k,t'}\rightarrow \delta_k }(\tilde{x}_{k,t'}=e^{j\omega}x_{i,t'}) \notag\\
&~~~\times P_{\tilde{x}_{k,t} \rightarrow \delta_k}(\tilde{x}_{k,t}=c)\notag\\
&= \frac{1}{C}\sum_{\substack{i\in \mathcal{I}_K, \omega\in \Omega\\ \text{for~} e^{j\omega}x_{i,t}=c}}
P_{\delta_{\mathbf{\Pi}} \rightarrow \boldsymbol{\pi}_k} (\boldsymbol{\pi}_k=\mathbf{e}_i)\notag\\
&~~~\times \prod_{t'=1}^{T_{\rm P}} P_{ \tilde{x}_{k,t'}\rightarrow \delta_k }(\tilde{x}_{k,t'}=e^{j\omega}x_{i,t'}),\quad c \in \mathcal{C}.
\label{MP.x}
\end{align}

The estimates of $\tilde{\mathbf{H}}$ and $\tilde{\mathbf{X}}$ from the message passing iteration contain phase and permutation ambiguities. We now describe how to estimate the phase and permutation ambiguities.
Specifically, the marginal posterior of $\sigma_k$ can be depicted as
\begin{align}
&\Delta_{\sigma_k} (\sigma_k)\notag\\
&\varpropto \Delta_{\delta_k \rightarrow \sigma_k} (\sigma_k)p_{\sigma_k}(\sigma_k)  \notag\\
&\varpropto \int_{\boldsymbol{\pi}_k, \{x_{k,t}\}_{t=1}^{T_{\rm P}}}\delta(\tilde{\mathbf{x}}_{{\rm P},k}-\sigma_k \boldsymbol{\pi}_k^{\rm T} {\mathbf{X}}_{\rm P})
\prod_{t=1}^{T_{\rm P}}  \Delta_{\tilde{x}_{k,t} \rightarrow \delta_k}(\tilde{x}_{k,t})\notag\\
&~~~~~~~~~~~~~~~~~~~~ \times \Delta_{\boldsymbol{\pi}_k \rightarrow \delta_k} (\boldsymbol{\pi}_k).
\end{align}
The discrete form of the message above can be written as
\begin{align}
&P_{\sigma_k} (\sigma_k = e^{j\omega})\notag\\ \label{MP.phase}
&= \frac{1}{C}\sum_{i\in \mathcal{I}_K} P_{\delta_{\mathbf{\Pi}} \rightarrow \boldsymbol{\pi}_k} (\boldsymbol{\pi}_k=\mathbf{e}_i)  \notag\\
&~~~~\times \prod_{t=1}^{T_{\rm P}}  P_{\tilde{x}_{k,t} \rightarrow \delta_k}(\tilde{x}_{k,t}=e^{j\omega}x_{i,t}),\quad \omega \in \Omega.
\end{align}
Then, an estimate of $\mathbf{\Sigma}$ is given by $\hat{\mathbf{\Sigma}}=\diag\{ \hat{\sigma}_1, \hat{\sigma}_2, \ldots,\hat{\sigma}_K\}$,
where $\hat{\sigma_k}= \arg\max_{ \omega \in \Omega} P_{\sigma_k} (e^{j\omega}) $.
The marginal posterior of $\mathbf{P}$ can be depicted as
\begin{align}
&\Delta_{\mathbf{P}} (\mathbf{P}) \notag\\
&\varpropto \Delta_{\delta_{\mathbf{\Pi}} \rightarrow \mathbf{P}} (\mathbf{P})p_{\mathbf{P}}(\mathbf{P})  \notag\\
& \varpropto \int_{\{\boldsymbol{\pi}_k\}_{k=1}^K} \delta([\boldsymbol{\pi}_1,\boldsymbol{\pi}_2,\ldots,\boldsymbol{\pi}_K]^{\rm T}-\mathbf{P})
\prod_{k=1}^K \Delta_{ \boldsymbol{\pi}_{k} \rightarrow \delta_{\mathbf{\Pi}}} (\boldsymbol{\pi}_{k}). \label{MP.p}
\end{align}
We write \eqref{MP.p} in its discrete form as
\begin{align}
&P_{\mathbf{P}} (\mathbf{P}=\mathbf{P}_\ell)=\frac{1}{C} \prod_{k=1}^K P_{ \boldsymbol{\pi}_{k} \rightarrow \delta_{\mathbf{\Pi}}} (\boldsymbol{\pi}_{k}=\mathbf{p}_{\ell,k}), \quad \ell \in \mathcal{I}_{K!}. \label{MP.permu}
\end{align}
Then, we obtain an estimate of $\mathbf{\Pi}$ by $\hat{\mathbf{\Pi}}=\arg\max P_{\mathbf{P}} (\mathbf{P}_\ell)$.

\begin{algorithm}[!htbp]
\xiaowuhao
\caption{\textbf{\!:} The SCSE algorithm}
\label{algorithm1}
\begin{algorithmic}[1]
 \REQUIRE $\mathbf{Y}$, prior distribution
 $p_{\tilde{\mathbf{H}}}(\tilde{\mathbf{H}})$,
 $p_{\tilde{\mathbf{X}}}(\tilde{\mathbf{X}})$, and
 $p_{\mathbf{Y}|\mathbf{Z}}(\mathbf{Y}|\mathbf{Z})$  \\
 \hspace{-0.55cm}$\mathbf{Initialization:}$ $\forall n,k,t$, $\hat{\tilde{h}}_{n,k}(1) = 0,v_{n,k}^{\tilde{h}} = 1, \hat{\tilde{x}}_{k,t}(1)$ is \notag\\
 \hspace{-0.55cm} randomly drawn from $\mathcal{C}$, $v_{k,t}^{\tilde{x}}(1)=1$, and $\hat{s}_{n,t}(0)=0 $\\
 \hspace{-0.55cm}$\mathbf{for}$ $m=1,\ldots,M_{max}$~~~~~~\%outer iteration  \\
 \hspace{-0.3cm}$ \mathbf{for}$ $l=1,\ldots,L_{max}$~~~~~~\%inter iteration \\
\hspace{-0.55cm}\% Message passing for part I \\
\STATE$\forall n,t$: $\bar{v}_{n,t}^p(l)=\sum_{k=1}^{K}|\hat{\tilde{h}}_{n,k}(l)|^2v_{k,t}^{\tilde{x}}(l)
+v_{n,k}^{\tilde{h}}(l)|\hat{\tilde{x}}_{k,t}(l)|^2$ \\
\STATE$\forall n,t$: $\bar{p}_{n,t}(l)=\sum_{k=1}^{K}\hat{\tilde{h}}_{n,k}(l)\hat{\tilde{x}}_{k,t}(l)$ \\
\STATE$\forall n,t$: $v_{n,t}^p(l)=\bar{v}_{n,t}^p(l)+\sum_{k=1}^{K}v_{n,k}^{\tilde{h}}(l)v_{k,t}^{\tilde{x}}(l)$ \\
\STATE$\forall n,t$: $\hat{p}_{n,t}(l)=\bar{p}_{n,t}(l)-\hat{s}_{n,t}(l-1)\bar{v}_{n,t}^p(l)$ \\
\STATE$\forall n,t$: $v_{n,t}^z(l)=\frac{v_{n,t}^p(l)\sigma^2}{v_{n,t}^p(l)+\sigma^2}$ \\
\STATE$\forall n,t$: $\hat{z}_{n,t}(l)=\frac{v_{n,t}^p(l)\sigma^2}{v_{n,t}^p(l)
+\sigma^2}(y_{n,t}-\hat{p}_{n,t}(l))+\hat{p}_{n,t}(l)$  \\
\STATE$\forall n,t$: $v_{n,t}^s(l)=(1-v_{n,t}^z(l)/v_{n,t}^p(l))/v_{n,t}^p(l)$\\
\STATE$\forall n,t$: $\hat{s}_{n,t}(l)=(\hat{z}_{n,t}(l)-\hat{p}_{n,t}(l))/v_{n,t}^p(l)$ \\
\STATE$\forall n,k$: $v_{n,k}^q(l)=(\sum_{t=1}^{T}|\hat{\tilde{x}}_{k,t}(l)|^2 v_{n,t}^s(l))^{-1}$ \\
\STATE$\forall n,k$: $\hat{q}_{n,k}(l)=\hat{\tilde{h}}_{n,k}(l)(1-\sum_{t=1}^{T}v_{k,t}^{\tilde{x}}(l) v_{n,t}^s(l))$ \notag\\
$~~~~~~~~~~~~~~~~~~~~+v_{n,t}^q(l)\sum_{t=1}^{T}(\hat{\tilde{x}}_{k,t}(l))^\ast\hat{s}_{n,t}(l)$\\
\STATE$\forall k,t$: $v_{k,t}^r(l)=(\sum_{n=1}^{N}|\hat{\tilde{h}}_{n,k}(l)|^2 v_{n,t}^s(l))^{-1}$ \\
\STATE$\forall k,t$: $\hat{r}_{k,t}(l)=\hat{\tilde{x}}_{k,t}(l)(1-\sum_{n=1}^{N}v_{n,k}^{\tilde{h}}(l) v_{k,t}^r(l))$\notag\\
$~~~~~~~~~~~~~~~~~~~~+v_{k,t}^r(l)\sum_{n=1}^{N}(\hat{\tilde{h}}_{n,k}(l))^\ast\hat{s}_{n,t}(l)$ \\
\STATE$\forall n,k$: $\hat{\tilde{h}}_{n,k}(l+1)= \mathbb{E}[\tilde{h}_{n,k}|\hat{q}_{n,k}(l), v_{n,k}^q(l)]$ \\
\STATE$\forall n,k$: $v_{n,k}^{\tilde{h}}(l+1)=\mathbb{E}[|\tilde{h}_{n,k}-\hat{\tilde{h}}_{n,k}(l+1)|^2|\hat{q}_{n,k}(l), v_{n,k}^q(l)]$ \\
\STATE$\forall k,t\in\mathcal{T}_{\rm D}$: $\hat{\tilde{x}}_{k,t}(l+1)=\mathbb{E}[\tilde{x}_{k,t}|\hat{r}_{k,t}(l), v_{k,t}^r(l)]$ \\
\STATE$\forall k,t\in\mathcal{T}_{\rm D}$: $v_{k,t}^{\tilde{x}}(l+1)=\mathbb{E}[|\tilde{x}_{k,t}-\hat{\tilde{x}}_{k,t}(l+1)|^2|\hat{r}_{k,t}(l), v_{k,t}^r(l)]$ \\
\hspace{-0.55cm}\% Message passing for part II \\
\STATE$\forall k,t\in\mathcal{T}_{\rm P}$: $P_{\tilde{x}_{k,t} \rightarrow \delta_k}^l(\tilde{x}_{k,t}=c) = \frac{1}{C} P_{\tilde{x}_{k,t}}(\tilde{x}_{k,t}=c)$\notag\\
$~~~~~~~~~\times \mathcal{CN}(\tilde{x}_{k,t}=c;\hat{r}_{k,t}(l),v_{k,t}^r(l)),\quad c\in \mathcal{C}$ \quad\% Eq.\eqref{MP.1}\\
\STATE$\forall k$: $p(\boldsymbol{\pi}_k=\mathbf{e}_i,\boldsymbol{\sigma}_k=\omega) $ \notag\\
$~~~~~= \frac{1}{C} \prod_{t=1}^{T_{\rm P}}\left[P_{ \tilde{x}_{k,t}\rightarrow \delta_k }^l(\tilde{x}_{k,t}= e^{j\omega} x_{i,t})\right],i \in \mathcal{I}_K,\omega \in \Omega $ \\
\STATE$\forall k$: $P_{\boldsymbol{\pi}_k \rightarrow \delta_{\mathbf{\Pi}}}^l (\boldsymbol{\pi}_k=\mathbf{e}_i)$ \notag\\
$~~~~~= \frac{1}{C} \sum_{\omega\in \Omega} p(\boldsymbol{\pi}_k=\mathbf{e}_i,\boldsymbol{\sigma}_k=\omega)$ \quad \% Eq.\eqref{MP.5},\eqref{MP.4} \\
\STATE$\forall k$: $P_{\delta_{\mathbf{\Pi}} \rightarrow \boldsymbol{\pi}_k}^l (\boldsymbol{\pi}_k=\mathbf{e}_i) = \frac{1}{C} \sum_{\substack{\ell=1\\ \ell:~ \mathbf{p}_{\ell,k}=\mathbf{e}_i }}^{K!}  $\notag\\
$\prod_{k'=1 \neq k}^K\left[P_{ \boldsymbol{\pi}_{k'} \rightarrow \delta_{\mathbf{\Pi}}}(\boldsymbol{\pi}_{k'}=\mathbf{p}_{\ell,k'})\right], \quad i \in \mathcal{I}_K  $\quad  \% Eq.\eqref{MP.8},\eqref{MP.7} \\
\STATE$\forall k,t\in\mathcal{T}_{\rm P}$: $P_{\tilde{x}_{k,t}}^{l+1} (\tilde{x}_{k,t}=c)=\frac{1}{C} \sum_{\substack{i \in \mathcal{I}_K, \omega \in \Omega \\ e^{j\omega} x_{i,t}=c}} P_{\delta_{\mathbf{\Pi}} \rightarrow \boldsymbol{\pi}_k}^l (\boldsymbol{\pi}_k=\mathbf{e}_i) $
$ p(\boldsymbol{\pi}_k=\mathbf{e}_i,\boldsymbol{\sigma}_k=\omega),\quad c\in \mathcal{C}$
\quad \% Eq.\eqref{MP.x} \\
\STATE$\forall k,t\in\mathcal{T}_{\rm P}$: $\hat{\tilde{x}}_{k,t}(l+1)=\mathbb{E}[x_{k,t}|\hat{r}_{k,t}(l), v_{k,t}^r(l), \mathbf{X}_{\rm P}]$ \\
\STATE$\forall k,t\in\mathcal{T}_{\rm P}$: $v_{k,t}^{\tilde{x}}(l+1)=$\notag\\
$~~~~~~~~~~~~\mathbb{E}[|x_{k,t}-\hat{\tilde{x}}_{k,t}(l+1)|^2|\hat{r}_{k,t}(l), v_{k,t}^r(l), \mathbf{X}_{\rm P}]$ \\
\STATE $\mathbf{if}\sum_{n,t}|\bar{p}_{n,t}(l)-\bar{p}_{n,t}(l-1)|^2 \leq \epsilon\sum_{n,t}|\bar{p}_{n,t}(l)|^2, \mathbf{stop}$ \\
\hspace{-0.3cm}$\mathbf{end}$ ~~~\\
\hspace{-0.3cm}$\forall k,t$: $\hat{\tilde{x}}_{k,t}(1)=\hat{\tilde{x}}_{k,t}(l+1)$;
$v_{k,t}^{\tilde{x}}(l)=v_{k,t}^{\tilde{x}}(l+1)$; \notag\\
\hspace{-0.3cm}$\forall n,k$: $\hat{\tilde{h}}_{n,k}(1)=0$; $v_{n,k}^{\tilde{h}}=1$~~~~~~\% Re-initialization  \\
\hspace{-0.55cm}$\mathbf{end}$ \\
\hspace{-0.55cm}\%Eliminate ambiguities\\
\STATE $\forall k$: $P_{\sigma_k} (\sigma_k = e^{j\omega})=\frac{1}{C}\sum_{i\in \mathcal{I}_K} P_{\delta_{\mathbf{\Pi}} \rightarrow \boldsymbol{\pi}_k}^{L_{max}} (\boldsymbol{\pi}_k=\mathbf{e}_i) $ \notag\\
$~~~~~\times \prod_{t=1}^{T_{\rm P}} P_{\tilde{x}_{k,t} \rightarrow \delta_k}^{L_{max}}(\tilde{x}_{k,t}=e^{j\omega}x_{i,t}),\quad \omega \in \Omega $ \quad \%Eq.\eqref{MP.phase} \\
\STATE $\forall k$: $\hat{\sigma}_k=\arg\max_{\omega\in\Omega}P_{\sigma_k} ( e^{j\omega}) $,
$\hat{\mathbf{\Sigma}}=\{ \hat{\sigma}_1, \hat{\sigma}_2, \ldots,\hat{\sigma}_K\}$ \\
\STATE $P_{\mathbf{P}} (\mathbf{P}=\mathbf{P}_\ell)=\frac{1}{C} \prod_{k=1}^K P_{ \boldsymbol{\pi}_{k} \rightarrow \delta_{\mathbf{\Pi}}}^{L_{max}} (\boldsymbol{\pi}_{k}=\mathbf{p}_{\ell,k}), \quad \ell \in \mathcal{I}_{K!}$ \\
~~~~~~~~~~~~~~~~~~~~~~~~~~~~~~~~~~~~~~~~~~~~~~~~~~~\%Eq.\eqref{MP.permu}\\
\STATE $\hat{\mathbf{\Pi}}= \arg\max_{\ell \in \mathcal{I}_{K!}} P_{\mathbf{P}}(\mathbf{P}_\ell) $\\
\ENSURE : $\hat{\mathbf{H}}=\tilde{\mathbf{H}}\hat{\mathbf{\Sigma}}\hat{\mathbf{\Pi}}$,
$\hat{\mathbf{X}}_{\rm D}=\hat{\mathbf{\Pi}}^{\rm -1}\hat{\mathbf{\Sigma}}^{\rm -1}\tilde{\mathbf{X}}_{\rm D}$  \\
\end{algorithmic}
\end{algorithm}

We are now ready to present the overall message passing algorithm in Algorithm~\ref{algorithm1}.
In specific,
 steps $1$ and $2$ update the estimate of the variance $\{\bar{v}_{n,t}^p(l)\}$ and the mean $\{\bar{p}_{n,t}(l)\}$ of  $\{z_{n,t}\}$, where $l$ denotes the inner iteration number. Steps $3$ and $4$ update the estimate of the variance $\{v_{n,t}^p(l)\}$ and the mean $\{\hat{p}_{n,t}(l)\}$ of $\{z_{n,t}\}$ by the ``Onsager" correction \cite{b17}.
Steps $5$ and $6$ give the estimate of the marginal posterior variance $\{v_{n,t}^z(l)\}$ and mean $\{\hat{z}_{n,t}(l)\}$ of $\{z_{n,t}\}$. Steps $7$ and $8$ calculate the inverse-residual-variances $\{v_{n,t}^s(l)\}$ and the scaled residual $\{\hat{s}_{n,t}(l)\}$. Steps $9$ and $10$ update the estimate of the variance $\{v_{n,k}^q(l)\}$ and mean $\{\hat{q}_{n,k}(l)\}$ of $\{\tilde{h}_{n,k}\}$ based on the messages from check nodes $\{p_{y_{n,t}|z_{n,t}}\}$.
Then, steps $13$ and $14$ give the estimate of the marginal posterior of $\{h_{n,k}\}$ with the mean $\{\hat{h}_{n,k}(l+1)\}$ and the variance $\{v_{n,k}^h(l+1)\}$.
Steps $11$ and $12$ process the same operations as steps $9$ and $10$ and steps $15$ and $16$ process the same operations as steps $13$ and $14$ for $\{\tilde{x}_{k,t}\in\mathbf{X}_{\rm D}\}$.
Step $17$ updates the messages from $\{\tilde{x}_{k,t}\}$ to $\{\delta_k\}$, where $\mathcal{CN}(\cdot;\hat{r}_{k,t}(l),v_{k,t}^r(l))$ is a Gaussian distribution obtained by $\prod_{n=1}^N \Delta_{p_{n,t}\rightarrow \tilde{x}_{k,t}}(\tilde{x}_{k,t})$.
Step $18$ calculate the probability of $\{\boldsymbol{\pi}_k = \mathbf{e}_i \}$ with phase shift $\{e^{j\omega}\}$.
Steps $19$ and $20$ update the messages between $\boldsymbol{\pi}_k$ and $\mathbf{\Pi}$.
Step $21$ updates the marginal posterior of $\{\tilde{x}_{k,t}\}$ by merging the messages from $\{p_{y_{n,t}|z_{n,t}}\}$
and $\{\delta_k\}$ with the priors $\{p_{\tilde{x}_{k,t}}(\tilde{x}_{k,t})\}$.
Steps $22$ and $23$ give the posterior mean and variance
of $\{\tilde{x}_{k,t}\}\in \mathbf{X}_{\rm P}$, where the expectations are taken over the distribution given by step $19$.
Step $24$ gives a stopping condition based on the (normalized) change of $\bar{p}_{n,t}(l)$ in two consecutive iteration and a user-defined parameter $\epsilon$.
Steps $25$ and $27$ give the posterior message of $\{\sigma_k\}$ and $\{\boldsymbol{\pi}_k\}$. Steps $26$ and $28$ give estimates of $\mathbf{\Sigma}$ and $\mathbf{\Pi}$.
Note that steps $25$-$28$ are out of the loop since the estimates of the phase and permutation ambiguities are not needed in the iterative process but are needed in correcting the estimates of $\mathbf{H}$ and $\mathbf{X}$ after message passing.

It is interesting to compare the joint channel-and-signal estimation scheme in \cite{ShiJin} with SCSE in Algorithm~\ref{algorithm1}.
In fact, Algorithm~\ref{algorithm1} can be modified for the joint channel-and-signal scheme as follows: i) In initialization, set $\hat{\tilde{x}}_{k,t}(1)=x_{k,t}$, $\forall k,t\in \mathcal{T}_{\rm P}$; ii) delete steps $17$-$23$, and $25$-$28$. That is, the main difference of SCSE from the joint channel-and-signal estimation scheme is that the former includes the estimation of the phase and permutation ambiguities, i.e., $\mathbf{\Sigma}$, and  $\mathbf{\Pi}$, in the iterative message passing process. We will show by numerical simulations that SCSE is able to significantly outperform the joint channel-and-signal estimation scheme in the short-pilot regime.

\subsection{S-SCSE Algorithm }
\label{Sec.S-SCSE}
\begin{figure}
  \centering
  \includegraphics[width=3 in]{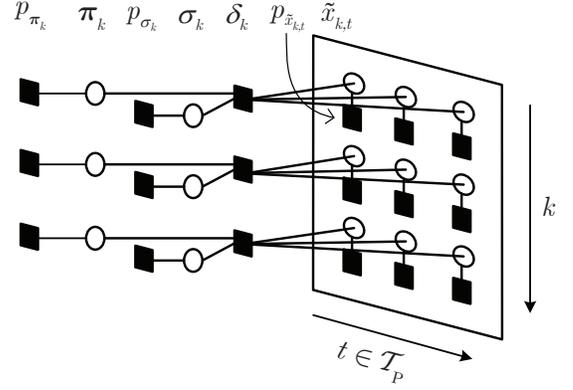}\\
  \caption{The simplified factor graph representation for Part II in Fig.~\ref{figure_factor} with  $K=3$ and $T_{\rm P}=3$.}\label{figure_factorII}
\end{figure}
The SCSE algorithm is computationally infeasible for a relatively large $k$, since it involves enumeration over all length-$k$ permutations in step $20$ of Algorithm~\ref{algorithm1}.
 To reduce the complexity, we relax the constraint that $\mathbf{\Pi}$ is a permutation to the one that each row $k$ of $\mathbf{\Pi}$ (denoted by $\boldsymbol{\pi}_k^{\rm T}$) is independently taken from the set $\{ \mathbf{e}_\ell\}_{\ell=1}^K$. That is,
\begin{align}
p_{\mathbf{\Pi}}(\mathbf{\Pi})
=p_{\mathbf{\Pi}}(\boldsymbol{\pi}_1,\boldsymbol{\pi}_2,\ldots,\boldsymbol{\pi}_K)
\thickapprox \prod_{k=1}^K p_{\boldsymbol{\pi}_k}(\boldsymbol{\pi}_k) \label{pi.1}
\end{align}
where $p_{\boldsymbol{\pi}_k}(\boldsymbol{\pi}_k)=1/K$, and $\boldsymbol{\pi}_k \in \{ \mathbf{e}_1, \mathbf{e}_2, \ldots, \mathbf{e}_K\}$.
The corresponding factor graph of part II in Fig.~\ref{figure_factor} is given in Fig.~\ref{figure_factorII}. Compared with part II in Fig.~\ref{figure_factor}, the factor graph in Fig.~\ref{figure_factorII} is almost the same, except that the nodes $\{p_{\mathbf{P}},\mathbf{P}, \sigma_{\mathbf{\Pi}}\}$ are replaced by
$\{ p_{\boldsymbol{\pi}_k}\}$.

We now describe message passing over the factor graph in Fig.~\ref{figure_factorII}.
The messages from $\tilde{x}_{k,t}$ to $\delta_k$ and from $\sigma_k$ to $\delta_k$ have the same form as in \eqref{MP.1} and \eqref{MP.2}. The message from $\boldsymbol{\pi}_k$ to $\delta_k$ is given by
\begin{align}
\Delta_{\boldsymbol{\pi}_k \rightarrow \delta_k}(\boldsymbol{\pi}_k)= p_{\boldsymbol{\pi}_k}(\boldsymbol{\pi}_k)
= \frac{1}{K}\sum_{i=1}^K \delta(\boldsymbol{\pi}_k = \mathbf{e}_i). \label{pi.2}
\end{align}
Then, the message from $\delta_k$ to $\tilde{x}_{k,t}$ is given by
\begin{align}
&\Delta_{\delta_k \rightarrow \tilde{x}_{k,t}}(\tilde{x}_{k,t}) \notag\\
&\varpropto \int_{\sigma_k, \boldsymbol{\pi}_k,  \{x_{k,t'}\}_{t'=1\neq t}^{T_{\rm P}} } \delta(\tilde{\mathbf{x}}_{{\rm P},k}-\sigma_k \boldsymbol{\pi}_k^{\rm T} {\mathbf{X}}_{\rm P})\notag\\
&~~~~~~~~\times p_{\boldsymbol{\pi}_k}(\boldsymbol{\pi}_k) p_{\sigma_k}(\sigma_k)  \prod_{t'=1 \neq t}^{T_{\rm P}} \Delta_{ \tilde{x}_{k,t'}\rightarrow \delta_k }(\tilde{x}_{k,t'}).  \label{pi.3}
\end{align}
Since $\tilde{x}_{k,t}$, $\boldsymbol{\pi}_k$, and $\sigma_k$ are discrete variables, we can write the above message in its discrete form as
\begin{align}
&P_{\delta_k \rightarrow \tilde{x}_{k,t}}(\tilde{x}_{k,t}=c) \notag\\
&= \frac{1}{C} \sum_{\substack{i \in \mathcal{I}_K, \omega \in \Omega \\ \text{for:~} e^{j\omega} x_{i,t}=c}} \prod_{t'=1\neq t}^{T_{\rm P}} P_{ \tilde{x}_{k,t'}\rightarrow \delta_k }(\tilde{x}_{k,t'}= e^{j\omega} x_{i,t'}),\notag\\
&~~~~~~~~~~~~~~~~~~~~~~~~~~~~~~~~~~~~~~~~~~~~~~~~~~~\quad c \in \mathcal{C}.  \label{pi.4}
\end{align}
Then, for any $c\in \mathcal{C}$, the marginal posterior of $\tilde{x}_{k,t}$ can be updated as
\begin{align}
&P_{\tilde{x}_{k,t}} (\tilde{x}_{k,t}=c)\notag\\
&\varpropto P_{\delta_k \rightarrow \tilde{x}_{k,t}}(\tilde{x}_{k,t}=c)
P_{\tilde{x}_{k,t} \rightarrow \delta_k}(\tilde{x}_{k,t}=c) \notag\\
&= \frac{1}{C} \sum_{\substack{i \in \mathcal{I}_K, \omega \in \Omega \\ \text{for:~} e^{j\omega} x_{i,t}=c}} \prod_{t'=1}^{T_{\rm P}} P_{ \tilde{x}_{k,t'}\rightarrow \delta_k }(\tilde{x}_{k,t'}= e^{j\omega} x_{i,t'}),\notag\\
&~~~~~~~~~~~~~~~~~~~~~~~~~~~~~~~~~~~~~~~~~~~~~~\quad c \in \mathcal{C}.
\label{pi.5}
\end{align}
The other messages are calculated by following the SCSE algorithm.
Compared to SCSE, S-SCSE omits the calculation of $P_{\delta_{\mathbf{\Pi}} \rightarrow \boldsymbol{\pi}_k} (\boldsymbol{\pi}_k=\mathbf{e}_i)$ in \eqref{MP.7}, which significantly reduces the computation complexity from $\mathcal{O}(K!)$ to $\mathcal{O}(K^2)$.

The phase and permutation ambiguities are estimated as follows.
The marginal posterior of $\sigma_k$ can be depicted as
\begin{align}
&\Delta_{\sigma_k} (\sigma_k)\notag\\
&\varpropto \Delta_{\delta_k \rightarrow \sigma_k} (\sigma_k)p_{\sigma_k}(\sigma_k)  \notag\\
&\varpropto \int_{\boldsymbol{\pi}_k, \{x_{k,t}\}_{t=1}^{T_{\rm P}}}\delta(\tilde{\mathbf{x}}_{{\rm P},k}-\sigma_k \boldsymbol{\pi}_k^{\rm T} {\mathbf{X}}_{\rm P})
\prod_{t=1}^{T_{\rm P}}  \Delta_{\tilde{x}_{k,t} \rightarrow \delta_k}(\tilde{x}_{k,t}). \label{Pi.6}
\end{align}
The discrete form of the above message can be written as
\begin{align}
&P_{\sigma_k} (\sigma_k = e^{j\omega})\notag\\
&= \frac{1}{C}  \sum_{ i \in \mathcal{I}_K}
\prod_{t=1}^{T_{\rm P}} P_{\tilde{x}_{k,t} \rightarrow \delta_k}(\tilde{x}_{k,t}= e^{j\omega}{x}_{i,t}),
\quad \omega \in \Omega\label{Pi.phase}.
\end{align}
Then, an estimate of $\mathbf{\Sigma}$ is given by $\hat{\mathbf{\Sigma}}=\diag\{ \hat{\sigma}_1, \hat{\sigma}_2, \ldots,\hat{\sigma}_K\}$,
where $\hat{\sigma_k}= \arg\max_{ \omega \in \Omega} P_{\sigma_k} (e^{j\omega}) $.
The marginal posterior of $\boldsymbol{\pi}_k$ can be depicted as
\begin{align}
&\Delta_{\boldsymbol{\pi}_k} (\boldsymbol{\pi}_k)\notag\\
&\varpropto \Delta_{\delta_k \rightarrow \boldsymbol{\pi}_k} (\boldsymbol{\pi}_k) p_{\boldsymbol{\pi}_k}(\boldsymbol{\pi}_k)  \notag\\
&\varpropto \int_{\sigma_k, \{x_{k,t}\}_{t=1}^{T_{\rm P}}} \delta(\tilde{\mathbf{x}}_{{\rm P},k}-\sigma_k \boldsymbol{\pi}_k^{\rm T} {\mathbf{X}}_{\rm P})
\prod_{t=1}^{T_{\rm P}}  \Delta_{\tilde{x}_{k,t} \rightarrow \delta_k}(\tilde{x}_{k,t}).
\end{align}
Similarly, the discrete form of the above message can be written as
\begin{align}
&P_{\boldsymbol{\pi}_k} (\boldsymbol{\pi}_k = \mathbf{e}_i)\notag\\
&= \frac{1}{C}  \sum_{\omega\in \Omega}
\prod_{t=1}^{T_{\rm P}} P_{\tilde{x}_{k,t} \rightarrow \delta_k}(\tilde{x}_{k,t}= e^{j\omega}{x}_{i,t}),
\quad i \in \mathcal{I}_K\label{Pi.permu}.
\end{align}
Then, an estimate of $\mathbf{\Pi}$ is given by
$\hat{\mathbf{\Pi}}=[ \hat{\boldsymbol{\pi}}_1, \hat{\boldsymbol{\pi}}_2, \ldots,\hat{\boldsymbol{\pi}}_K]^{\rm T}$,
where $\hat{\boldsymbol{\pi}}_k= \arg\max_{\ell \in \mathcal{I}_K} P_{\boldsymbol{\pi}_k} (\mathbf{e}_\ell) $.

\begin{algorithm}[t]
\xiaowuhao
\vspace{0.2mm}
\caption{\textbf{\!:} \vspace{0.15mm} The S-SCSE algorithm}
\label{algorithm2}
\begin{algorithmic}[1]
 \REQUIRE $\mathbf{Y}$, prior distribution
 $p_{\tilde{\mathbf{H}}}(\tilde{\mathbf{H}})$,
 $p_{\tilde{\mathbf{X}}}(\tilde{\mathbf{X}})$, and
 $p_{\mathbf{Y}|\mathbf{Z}}(\mathbf{Y}|\mathbf{Z})$  \\
 \hspace{-0.55cm}$\mathbf{Initialization:}$ $\forall n,k,t$, $\hat{\tilde{h}}_{n,k}(1) = 0,v_{n,k}^{\tilde{h}} = 1, \hat{\tilde{x}}_{k,t}(1)$ is \notag\\
 \hspace{-0.55cm} randomly drawn from $\mathcal{C}$, $v_{k,t}^{\tilde{x}}(1)=1$, and $\hat{s}_{n,t}(0)=0 $\\
 \hspace{-0.55cm}$\mathbf{for}$ $m=1,\ldots,M_{max}$~~~~~~\%outer iteration  \\
 \hspace{-0.3cm}$ \mathbf{for}$ $l=1,\ldots,L_{max}$~~~~~~\%inter iteration \\
 \% Message passing for part I \\
 Perform steps $1$-$16$ in Algorithm~\ref{algorithm1}. \\
 \% Message passing for part II \\
 Perform steps $17$-$18$ in Algorithm~\ref{algorithm1}. \\
\STATE $\forall (k,t)\in\mathbf{X}_{\rm P}$: $P_{\tilde{x}_{k,t}}^{l+1} (\tilde{x}_{k,t}=c)= \frac{1}{C} $ \notag\\
$ ~~~~~~~~~~~~\times\sum_{\substack{i \in \mathcal{I}_K, \omega \in \Omega \\ e^{j\omega} x_{i,t}=c}} p(\boldsymbol{\pi}_k=\mathbf{e}_i,\boldsymbol{\sigma}_k=\omega) \quad c\in \mathcal{C}$\\
~~~~~~~~~~~~~~~~~~~~~~~~~~~~~~~~~~~~~~~~~~~~~~~~~~~\% Eq. \eqref{pi.5} \\
\STATE $\forall k,t\in\mathcal{T}_{\rm P}$: $\hat{\tilde{x}}_{k,t}(l+1)=\mathbb{E}[x_{k,t}|\hat{r}_{k,t}(l), v_{k,t}^r(l), \mathbf{X}_{\rm P}]$ \\
\STATE$\forall k,t\in\mathcal{T}_{\rm P}$: $v_{k,t}^{\tilde{x}}(l+1) =$ \notag\\
$~~~~~~~~~~~~~~~~~\mathbb{E}[|x_{k,t}-\hat{\tilde{x}}_{k,t}(l+1)|^2|\hat{r}_{k,t}(l), v_{k,t}^r(l), \mathbf{X}_{\rm P}]$ \\
\STATE $\mathbf{if}\sum_{n,t}|\bar{p}_{n,t}(l)-\bar{p}_{n,t}(l-1)|^2 \leq \epsilon\sum_{n,t}|\bar{p}_{n,t}(l)|^2, \mathbf{stop}$ \\
\hspace{-0.3cm}$\mathbf{end}$ \\
\hspace{-0.3cm}$\forall k,t$: $~~~\hat{\tilde{x}}_{k,t}(1)=\hat{\tilde{x}}_{k,t}(l+1)$; $v_{k,t}^{\tilde{x}}(l)=v_{k,t}^{\tilde{x}}(l+1)$; \notag\\
\hspace{-0.3cm}$\forall n,k$: $\hat{\tilde{h}}_{n,k}(1)=0$; $v_{n,k}^{\tilde{h}}=1$ ~~~~~~\% Re-initialization  \\
\hspace{-0.55cm}$\mathbf{end}$ \\
\hspace{-0.55cm}\% Eliminate ambiguities \\
\STATE $\forall k$: $P_{\sigma_k} (\sigma_k = e^{j\omega})  $ \notag\\
$~=\frac{1}{C}\sum_{ i \in \mathcal{I}_K}
\prod_{t=1}^{T_{\rm P}} P_{\tilde{x}_{k,t} \rightarrow \delta_k}^{L_{max}}(\tilde{x}_{k,t}= e^{j\omega}{x}_{i,t}),\quad \omega\in \Omega $ \\
~~~~~~~~~~~~~~~~~~~~~~~~~~~~~~~~~~~~~~~~~~~~~~~~~~~~~~\% Eq. \eqref{Pi.phase}\\
\STATE $\forall k$: $P_{\boldsymbol{\pi}_k} (\boldsymbol{\pi}_k = \mathbf{e}_\ell)  $\notag\\
$~=\frac{1}{C}  \sum_{\omega \in \Omega}
\prod_{t=1}^{T_{\rm P}} P_{\tilde{x}_{k,t} \rightarrow \delta_k}^{L_{max}}(\tilde{x}_{k,t}= e^{j\omega}{x}_{i,t}),\quad \ell\in \mathcal{I}_K $ \\
~~~~~~~~~~~~~~~~~~~~~~~~~~~~~~~~~~~~~~~~~~~~~~~~~~~~~~\% Eq. \eqref{Pi.permu}\\
\STATE $\forall k$: $\hat{\sigma}_k=\arg\max_{\omega\in\Omega}P_{\sigma_k} ( e^{j\omega}) $,
$\hat{\mathbf{\Sigma}}=\{ \hat{\sigma}_1, \hat{\sigma}_2, \ldots,\hat{\sigma}_K\}$ \\
\STATE $\forall k$: $\hat{\boldsymbol{\pi}}_k=\arg\max_{\ell \in \mathcal{I}_K}P_{\boldsymbol{\pi}_k}(\mathbf{e}_\ell)$,
$\hat{\mathbf{\Pi}}=\{ \hat{\boldsymbol{\pi}}_1, \hat{\boldsymbol{\pi}}_2, \ldots,\hat{\boldsymbol{\pi}}_K\}$ \\
\ENSURE : $\hat{\mathbf{H}}=\tilde{\mathbf{H}}\hat{\mathbf{\Sigma}}\hat{\mathbf{\Pi}}$,
$\hat{\mathbf{X}}_{\rm D}=\hat{\mathbf{\Pi}}^{\rm -1}\hat{\mathbf{\Sigma}}^{\rm -1}\tilde{\mathbf{X}}_{\rm D}$  \\
\end{algorithmic}
\end{algorithm}

The S-SCSE algorithm is presented in Algorithm~\ref{algorithm2}. In Algorithm~\ref{algorithm2}, the S-SCSE algorithm performs the steps
the steps $1$-$18$ of Algorithm~\ref{algorithm1} first.
Then, step $1$ of Algorithm~\ref{algorithm2} updates the marginal posterior of $\{\tilde{x}_{k,t}\}$ by S-SCSE algorithm.
Steps $2$ and $3$ give the posterior mean and variance
of $\{\tilde{x}_{k,t}\}\in \mathbf{X}_{\rm P}$, where the expectations are taken over the distribution given by steps $1$.
Steps $5$ and $6$ give the posterior messages of $\sigma_k$ and $\boldsymbol{\pi}_k$ by S-SCSE algorithm, respectively.
Steps $7$ and $8$ gives estimates of $\mathbf{\Sigma}$ and $\mathbf{\Pi}$, respectively.\footnote{To identify users uniquely, $T_{\rm P}$ is required to be large enough to ensure that for each $\mathbf{x}_{{\rm P}, k}$ and $\mathbf{x}_{{\rm P}, k'}$, $\mathbf{x}_{{\rm P}, k} \neq e^{j\omega}\mathbf{x}_{{\rm P}, k'}$,
 $\omega \in \Omega$. This implies that, when the data are modulated by quaternary phase shift keying (QPSK), $T_{\rm P}$ should be no less than $1+\lceil\frac{1}{2}\log_2 K\rceil$, where one symbol is used to correct the phase shift of each user $k$, and $\lceil\frac{1}{2}\log_2 K\rceil$ symbols are used to guarantee that the pilot sequences of the $K$ users are different from each other.}
 In addition, the damping technique is used in the algorithm to improve convergence in simulation. We refer readers to \cite{b16} and \cite{liuhang} for more details.

\subsection{ Parameters Tuning}

\label{Para-tun}

Note that both the SCSE and S-SCSE algorithms require the knowledge of the distributions of $p_{\tilde{\mathbf{H}}}(\tilde{\mathbf{H}})$, $p_{\tilde{\mathbf{X}}}(\tilde{\mathbf{X}})$, and $p_{\mathbf{Y}|\mathbf{Z}}(\mathbf{Y}|\mathbf{Z})$ to perform the message passing process.
However, the parameters $\{\rho, \sigma_{h,k}^2, \forall k, N_0\}$ are usually difficult to acquire prior to the detection procedure\cite{b16}. Therefore, these model parameters need to be estimated as well.

In this paper, we use the expectation maximization (EM) algorithm\cite{dempster1977maximum} to tune these model parameters by taking $\tilde{\mathbf{H}}$ and $\tilde{\mathbf{X}}$ as the hidden variables. The specifical update rules are given as follows.
\begin{align}
&\rho^{(m+1)}
= \arg\max_{\rho} \mathbb{E} \left[ \log p \left(\tilde{\mathbf{H}},\tilde{\mathbf{X}},\mathbf{Y};\rho,(\sigma_{h,1}^2)^{(m)},\ldots,\right.\right. \notag\\
 &~~~~~~~~~~~\left.\left. (\sigma_{h,K}^2)^{(m)},N_0^{(m)} \right) \right].
\end{align}
\begin{align}
&(\sigma_{h,k}^2)^{(m+1)}
= \arg\max_{\sigma_{h,k}^2} \mathbb{E} \left[ \log p \left(\tilde{\mathbf{H}},\tilde{\mathbf{X}},\mathbf{Y};\rho^{(m+1)}, \right.\right. \notag\\
&~~~~~~~~~~~~~~~\left.\left. (\sigma_{h,1}^2)^{(m+1)},\ldots, (\sigma_{h,k-1}^2)^{(m+1)}, (\sigma_{h,k}^2), \right.\right. \notag\\
&~~~~~~~~~~~~~~~\left.\left.  (\sigma_{h,k+1}^2)^{(m)}, \ldots(\sigma_{h,K}^2)^{(m)} ,N_0^{(m)} \right) \right],  \forall k.
\end{align}
\begin{align}
&N_0^{(m+1)}
= \arg\max_{N_0} \mathbb{E} \left[ \log p \left(\tilde{\mathbf{H}},\tilde{\mathbf{X}},\mathbf{Y};\rho^{(m+1)},\right.\right. \notag\\
 &~~~~~~~~~~~~\left.\left. (\sigma_{h,1}^2)^{(m+1)},\ldots, (\sigma_{h,K}^2)^{(m+1)},N_0 \right) \right].
\end{align}
The EM update is performed in each outer iteration and the expectations in $(42)$-$(44)$ are taken over the approximate marginal posteriors
$\left\{p_{\tilde{h}_{n,k}|\mathbf{Y}},p_{\tilde{x}_{k,t}|\mathbf{Y}},p_{\tilde{z}_{n,t}|\mathbf{Y}} \right\}_{\forall n,k,t}$ obtained from every $L_{max}$th inner iteration.

\subsection{Complexity Analysis}

We now compare the computational complexity of our proposed algorithms with the existing approaches. Since both the joint channel-and-signal (JCSE) scheme in \cite{ShiJin} and the blind detection scheme in \cite{b15} are based on the BiG-AMP algorithm \cite{b16}, we only need to compare the computational complexity of the BiG-AMP, SCSE, and S-SCSE algorithms.

The computational complexity in steps $4$-$8$ of Algorithm~\ref{algorithm1} is $\mathcal{O}(NT)$, and that in steps $1$-$3$ and steps $9$-$16$ is $\mathcal{O}(NK + KT)$. Since the BiG-AMP algorithm only perform steps $1$-$16$ in Algorithm~\ref{algorithm1} for each iteration, the computational complexity of the BiG-AMP algorithm is $\mathcal{O}(NT)+ \mathcal{O}(NK + KT)$\cite{b16}. The computational complexity of steps $17$-$21$ in Algorithm~\ref{algorithm1} is
$\mathcal{O}(K T_{\rm P})$, $\mathcal{O}(K^2 T_{\rm P})$, $\mathcal{O}(K)$, $\mathcal{O}(K^2K!)$, and $\mathcal{O}(K^2 T_{\rm P})$, respectively. With the increase of $K$, $\mathcal{O}(K^2K!)$ dominates the complexity. Then, the overall computational complexity of the SCSE algorithm is $\mathcal{O}(NT)+ \mathcal{O}(NK + KT)+ \mathcal{O}(K^2K!)$ per iteration.
The computational complexity of step $1$ in Algorithm~\ref{algorithm2} is $\mathcal{O}(K^2 T_{\rm P})$. Thus, the computational complexity of the S-SCSE algorithm is
$\mathcal{O}(NT)+ \mathcal{O}(NK + KT)+ \mathcal{O}(K^2 T_{\rm P})$ per iteration. The computational complexity for our considered algorithms are summarized in Table~\ref{tab:Com}.

\begin{table}[htbp]
\scriptsize
\caption{Computational Complexity}
\centering
\begin{tabular}{l  l}
\toprule
  Method  & Complexity \\
  \hline
  JCSE   & $\mathcal{O}(NT) + \mathcal{O}(NK + KT)$\\
  Blind detection  & $\mathcal{O}(NT) + \mathcal{O}(NK + KT)$\\
  SCSE  & $\mathcal{O}(NT) + \mathcal{O}(NK + KT) + \mathcal{O}(K^2K!)$\\
  S-SCSE & $\mathcal{O}(NT)+ \mathcal{O}(NK + KT)+ \mathcal{O}(K^2 T_{\rm P})$\\
\bottomrule
\end{tabular}\label{tab:Com}
\end{table}

From Table~\ref{tab:Com}, we see that, compared to the JCSE and blind detection schemes, the complexity of the S-SCSE algorithm
is dominated by the third term $\mathcal{O}(K^2K!)$ caused by the estimation of $\mathbf{\Sigma}$ and $\mathbf{\Pi}$.
By relaxing the permutation constraint, the S-SCSE algorithm can significantly reduce the computational complexity of estimating $\mathbf{\Sigma}$ and $\mathbf{\Pi}$ from $\mathcal{O}(K^2K!)$ to $\mathcal{O}(K^2 T_{\rm P})$.

\subsection{Metric for Random Initializations}


The semi-blind detection problem in \eqref{Sys2} is non-convex, and the SCSE and S-SCSE algorithms are prone to be stuck at local optima. To alleviate this issue, multiple random initializations and multiple re-initializations are conducted.

We next describe how to choose a desirable result among multiple random initializations.
In a practical receiver, the metrics such as the mean-square error of the channel and the symbol error rate of the signal are not useful in evaluating the performance of random initializations since the ground truth is not available to the receiver. In this regard, we propose to use the
following heuristic metric for evaluating random initializations:
\begin{align}
&J(\tau)= \|\mathbf{Y} - \mathbf{H}(\tau)\mathbf{X}(\tau)\|_{\rm F}^2  \label{ja.1}
\end{align}
where $\tau$ is the index of random initializations. We choose the initialization with the minimum value of $J(\tau)$.

\section{Numerical Results}
\label{sec.Num}

In simulations, the signals are taken from quadrature phase shift keying (QPSK) or $16$ QAM with Gray-mapping.
We set $\alpha_k=1/K$ and $P=K$. The SNR is defined by $\frac{K}{N_0}$.
Following \cite{ShiJin}\cite{b15},
we divide all the AoA into $N$ grids and assume that all received signals from angle $e^{-j2\pi \frac{(n-1) d}{2\lambda}\cos(\theta_{\ell,k})}$ to angle $e^{-j2\pi \frac{n d}{2\lambda}\cos(\theta_{\ell,k})}$ belong to the $n$th grid. Thus,  the array steering matrix $\mathbf{A}_r$ can be regarded as a DFT matrix.
The aggregated channel gains are generated from the B-CSCG distribution in \eqref{channel.H}. The channel powers $\sigma_{h,k}^2, \forall k$ are randomly drawn from a uniform distribution over $[\sigma_{h,min}^2, 1]$.
 For the simulated algorithms, the maximum number of inner iterations $L_{\rm max}$ is set to $200$, and the
maximum number of outer iterations $M_{\rm max}$ is set to $10$.
The simulation results presented in this paper are obtained by taking average over 100 random realizations.
We compare the  numerical results of various approaches, as listed below.
\begin{figure}[t]
  \centering
  \includegraphics[width=3.7 in]{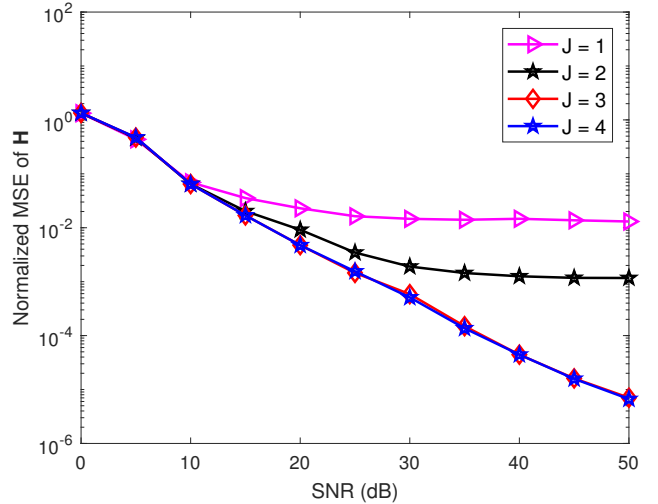}\\
  \caption{The normalized MSE of $\mathbf{H}$ versus SNR with the number of initializations ranging from $1$ to $4$ for the S-SCSE scheme with QPSK modulation. $K=20$, $N=200$, $\rho=0.2$, $T=50$, $T_{\rm P}=4$, and $\sigma_{h,min}^2=1$.}\label{Ini_times}
\end{figure}
\begin{itemize}
  \item { {OMP:}} A separate channel-and-signal detection approach to estimate $\mathbf{H}$ by orthogonal matching pursuit (OMP)\cite{tropp2007signal} with the pilots only.
  \item { {CoSaMP:}} A separate channel-and-signal detection approach to estimate $\mathbf{H}$ by compressive sampling matching pursuit (CoSaMP)\cite{needell2009cosamp} with the pilots only.
  \item { {Turbo-CS:}} A separate channel-and-signal detection approach to estimate $\mathbf{H}$ by turbo compressed sensing (Turbo-CS)\cite{chen2018structured} with the pilots only.
  \item { {JCSE:}} The joint channel-and-signal detection scheme in \cite{ShiJin}.
  \item {{BD:}} The blind detection scheme based on the BiG-AMP algorithm \cite{b16}.
  \item {{ SCSE:}} The SCSE algorithm proposed in this paper.
  \item {{ S-SCSE:}} The S-SCSE algorithm proposed in this paper.
  \item {{ LB-$\mathbf{H}$:}} To estimate $\mathbf{H}$ by the BiG-AMP algorithm in \cite{b16} with perfectly known $\mathbf{X}$.
  \item {{ LB-$\mathbf{X}$:}} To estimate $\mathbf{X}$ by the BiG-AMP algorithm in \cite{b16} with perfectly known positions of the non-zero elements of $\mathbf{H}$.
\end{itemize}

\begin{figure}[t]
  \centering
  \includegraphics[width=3.7 in]{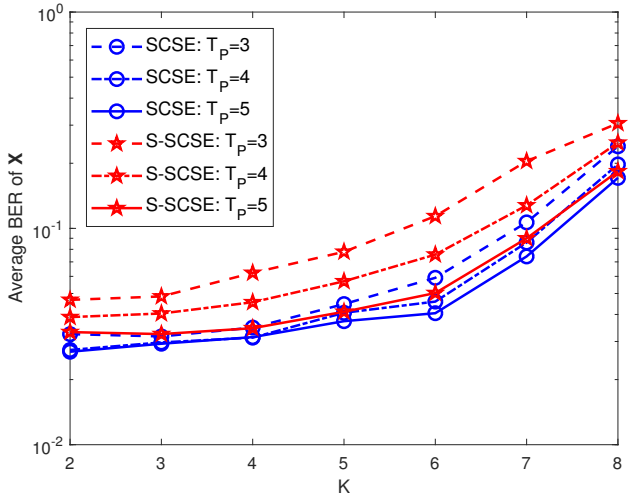}
  \caption{Comparison of the BER of $\mathbf{X}$ versus the number of users $K$ for SCSE and S-SCSE with the number of pilots
  $T_{\rm P}= 3$, $4$, and $5$, SNR $=0$ dB, $N/K=10$, $\rho=0.5$, $T=50$, and $\sigma_{h,min}^2=1$.}\label{S_SCSE}
\end{figure}
\begin{figure}[t]
  \centering
  \includegraphics[width=3.7 in]{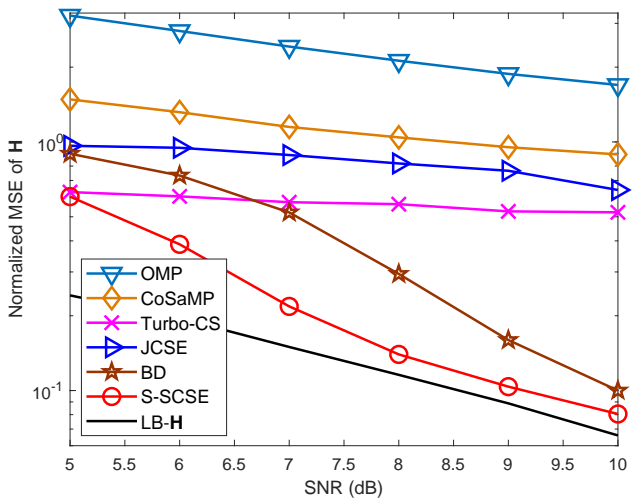}
  \caption{Comparison of the normalized MSE of $\mathbf{H}$ versus SNR for the OMP, CoSaMP, Turbo-CS,
   JCSE, BD, S-SCSE, and the LB-$\mathbf{X}$ with the pilots number $T_{\rm P}=8$, $K=20$, $N=128$, $\rho=0.3$, $T=50$, and $\sigma_{h,min}^2=1$.}\label{Channel}
\end{figure}
\begin{figure}[t]
  \centering
  \subfigure[QPSK]{\includegraphics[width=3.7 in]{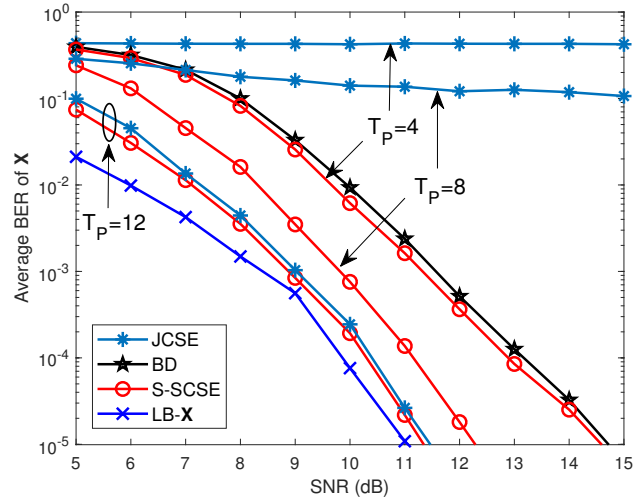}}
  \subfigure[16QAM]{\includegraphics[width=3.7 in]{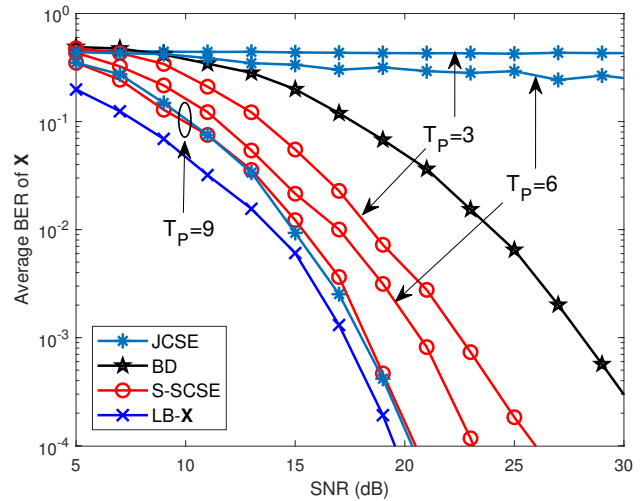}}
  \caption{The BER of $\mathbf{X}$ versus SNR for the JCSE, BD, S-SCSE, and LB-$\mathbf{X}$ with different numbers of pilots. $K=20$, $N=128$, $\rho=0.3$, $T=50$, and $\sigma_{h,min}^2=1$.}\label{X}
\end{figure}

Fig.~\ref{Ini_times} compares the normalized mean-square error (MSE) of $\mathbf{H}$
 versus SNR with different numbers of random initializations varying from $J=1$ to $J=4$ for the S-SCSE algorithm with QPSK modulation. The other settings are $K=20$, $N=200$, $\rho=0.2$, $T=50$, and $\sigma_{h,min}^2=1$.
We see that with the metric in \eqref{ja.1}, random initialization substantially improve the performance of the semi-blind scheme.

Fig.~\ref{S_SCSE} compares the average bit error rate (BER) of $\mathbf{X}$ versus the number of users $K$ for SCSE and S-SCSE with the number of pilots $T_{\rm P}=$ 3, 4, and 5. The other settings are SNR $=0$ dB, $N/K=10$, $\rho=0.5$, $T=50$, $\sigma_{h,min}^2=1$, and $J=5$. We can see that for a relatively large $T_{\rm P}$ (say, $T_{\rm P}=5$ for the configuration in Fig.~\ref{S_SCSE}), S-SCSE is able to perform close to SCSE. Note that due to high computational complexity for SCSE, we hence forth only present the simulation results of S-SCSE.

Fig.~\ref{Channel} compares the normalized mean-square error (MSE) of $\mathbf{H}$ versus SNR for the OMP, CoSaMP, Turbo-CS,
   JCSE, BD, S-SCSE, and LB-$\mathbf{X}$ with the pilots number $T_{\rm P}=8$.
 The other setting are $K=20$, $N=128$, $\rho=0.3$, $T=50$, $\sigma_{h,min}^2=1$, and $J=5$.
 From Fig.~\ref{Channel}, we see that our S-SCSE algorithm  significantly outperforms the training-based schemes (including OMP, CoSaMP, Turbo-CS, and JCSE) and  blind detection scheme. We also see that with the increase of SNR, the performance of our S-SCSE algorithm can approach that of the LB-$\mathbf{X}$.

\begin{figure}[t]
  \centering
  \includegraphics[width=3.7 in]{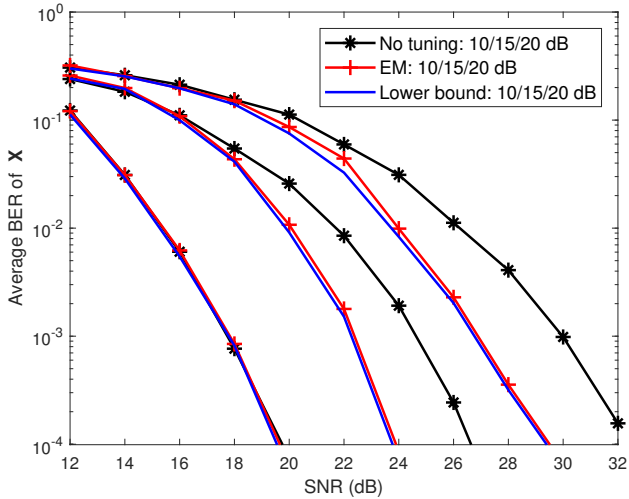}
  \caption{Comparison of the BER of $\mathbf{X}$ versus SNR for no tuning, EM, and known $\sigma_{h,k}$ with $-10\log_{10}(\sigma_{h,min}^2) = 10/15/20$ dB of the S-SCSE algorithm, $K=20$, $N=128$, $\rho=0.3$, and $T=50$.}\label{EM}
\end{figure}
\begin{figure}[t]
  \centering
  \includegraphics[width=3.7 in]{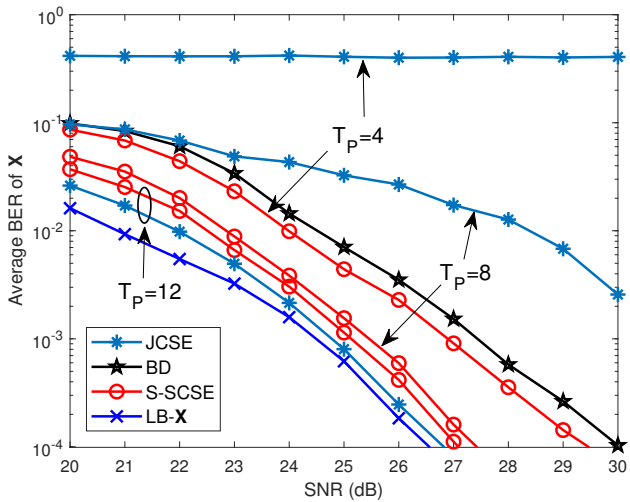}
  \caption{The BER of $\mathbf{X}$ versus SNR for the JCSE, BD, and S-SCSE with different numbers of pilots. $K=20$, $N=128$, $\rho=0.3$, $T=50$, and $-10\log_{10}(\sigma_{h,min}^2) = 20$ dB.}\label{Fading20}
\end{figure}
\begin{figure}[t]
  \centering
  \subfigure[$\rho = 0.1$]{\includegraphics[width=1.7 in]{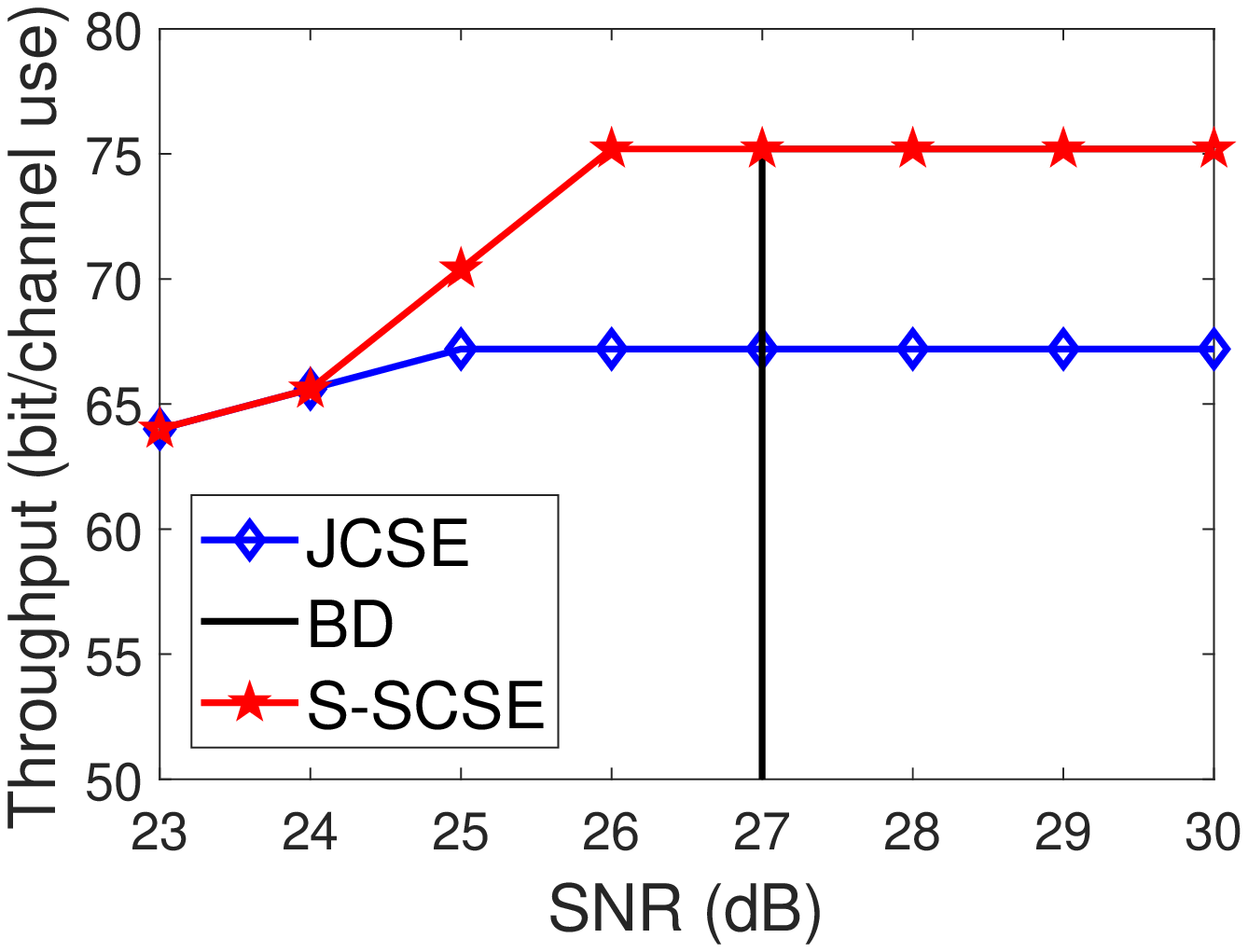}}
  \subfigure[$\rho = 0.2$]{\includegraphics[width=1.7 in]{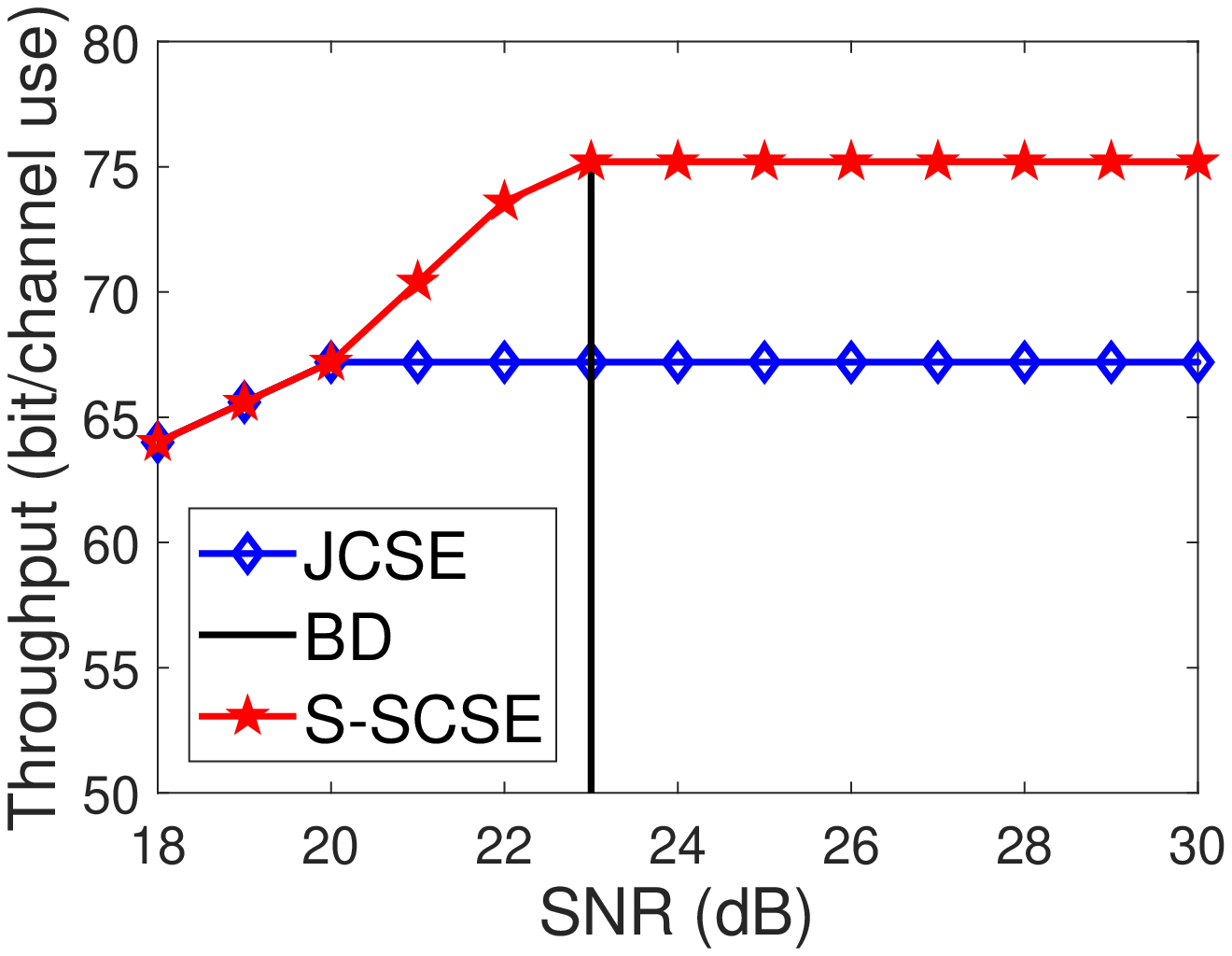}}
  \subfigure[$\rho = 0.3$]{\includegraphics[width=1.7 in]{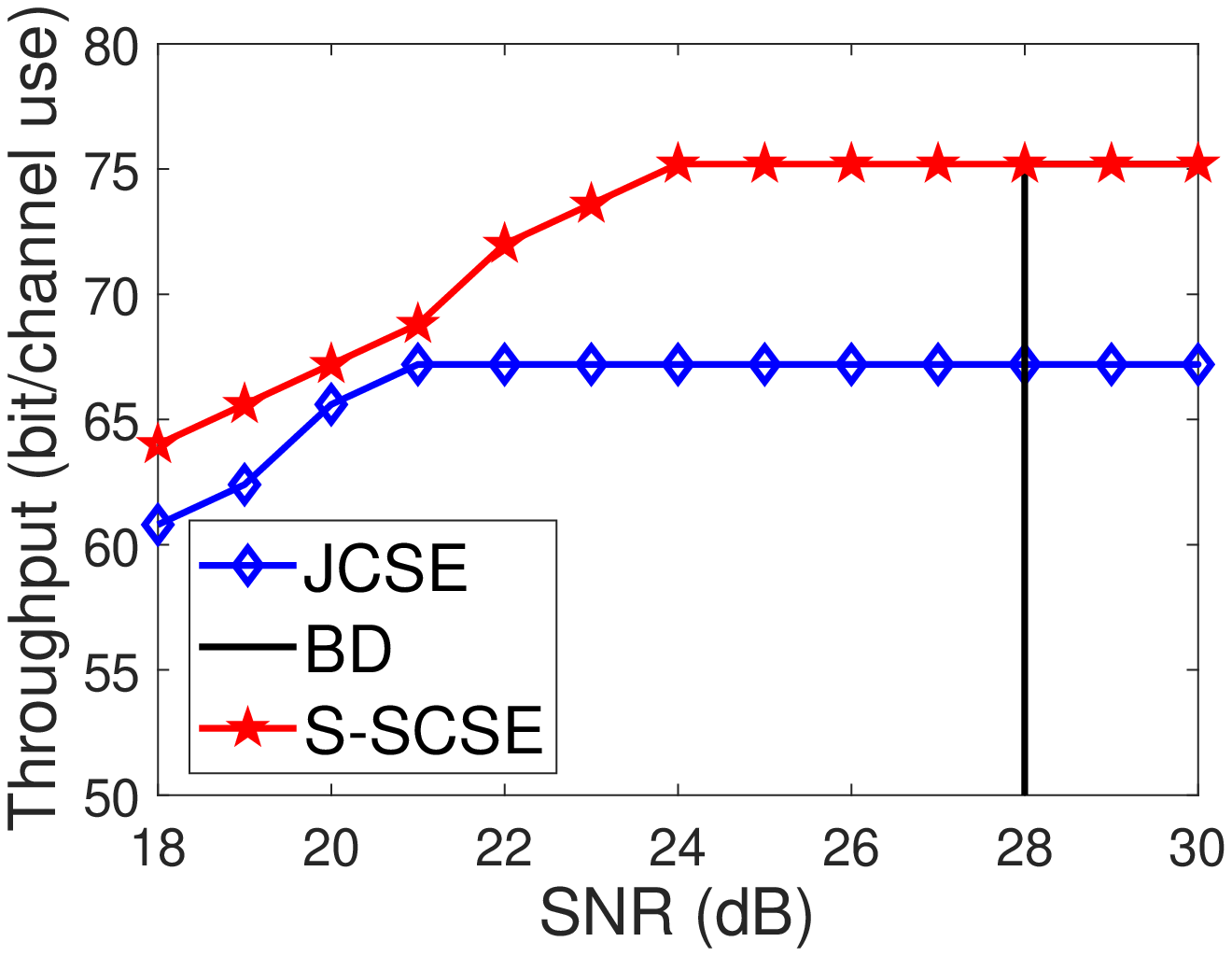}}
  \subfigure[$\rho = 0.4$]{\includegraphics[width=1.7 in]{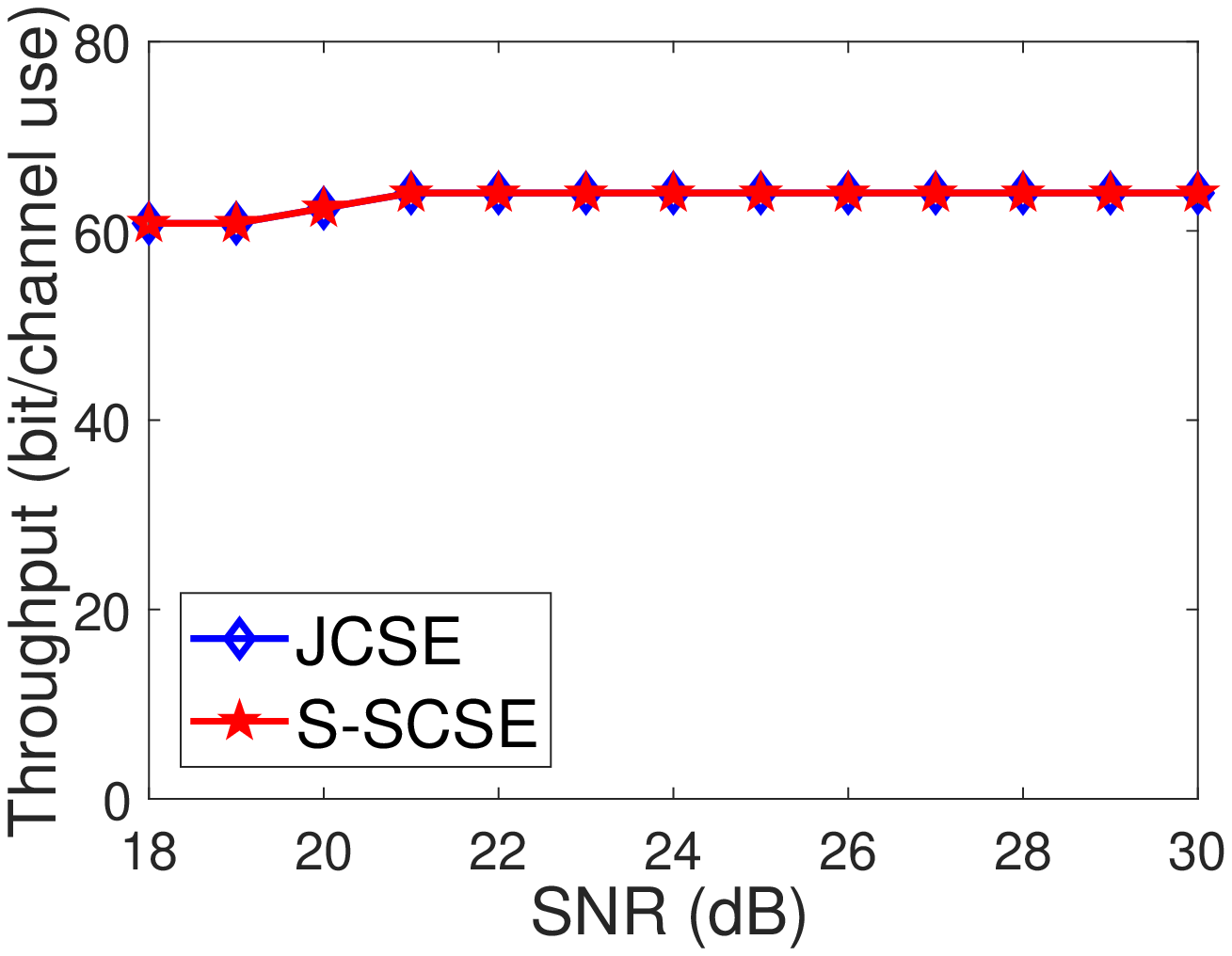}}
  \caption{Comparison of the throughput between different sparsity with 16 QAM modulation versus SNR for the JCSE, BD, and S-SCSE, $K=20$, $N=200$, $T=50$, and $\sigma_{h,min}^2=1$.}\label{throughput}
\end{figure}

Fig.~\ref{X} presents the bit error rate (BER) of $\mathbf{X}$ versus SNR for the JCSE, BD, S-SCSE, and LB-$\mathbf{X}$ with different numbers of pilot symbols. The other settings are $K=20$, $N=128$, $\rho=0.3$, $T=50$, and $\sigma_{h,min}^2=1$.
For Fig.~\ref{X}(a), we set $J=5$, while for Fig.~\ref{X}(a), we set $J=12$. Note that when the modulation is changed from QPSK (in Fig.~\ref{X}(a)) to
16-QAM (in Fig.~\ref{X}(b)), a lager number of random initializations is required to ensure stable semi-blind detection.
Also note that the blind detection system needs one reference symbol and a user label. For the simulation settings considered here, this amounts to a cost of $1+\lceil{\frac{1}{2}\log_2 K}\rceil=4$ symbols for QPSK modulation, and $1+\lceil{\frac{1}{2}\log_4 K}\rceil=3$ symbols for 16-QAM modulation.

In Fig.~\ref{X}(a), we see that for $T_{\rm P}=4$ and $8$, S-SCSE significantly outperforms the JCSE scheme. We also see that for
$T_{\rm P}=4$, S-SCSE slightly outperforms the blind detection scheme, while for $T_{\rm P}=8$, SCSE outperforms the blind detection scheme by about $4$ dB at BER $=10^{-5}$. For $T_{\rm P}=12$, the S-SCSE and JCSE schemes perform close to each other and approximate the lower bound. The reason is that in this case $T_{\rm P}$ is large enough to provide a relatively accurate initial channel estimate, and so the training-based scheme can work well.
In Fig.~\ref{X}(b), S-SCSE can outperforms the blind detection scheme by about $6$ dB at BER $=10^{-3}$ for $T_{\rm P}=3$,
and by about $10$ dB at BER $=10^{-4}$ for $T_{\rm P}=9$.  Similarly, S-SCSE significantly outperforms the JCSE scheme for $T_{\rm P}=3$ and $6$. The S-SCSE and JCSE schemes perform close to each other for $T_{\rm P}=12$.
 Fig.~\ref{X} shows that both the JCSE and S-SCSE schemes achieve better performance with the increase of $T_{\rm P}$.
 However, such performance improvement is achieved at the cost of a decrease in spectrum efficiency,
 since the pilots cannot transmit information. This issue will be elaborated later in Fig.~\ref{throughput}.

We next study the impact of large-scale fading on the system performance. In simulations, we set $-10\log_{10}(\sigma_{h,min}^2) = 10/15/20$ dB. The other setting are $K=20$, $N=128$, $\rho=0.3$, $T=50$, and $J=5$.
We consider the different configurations of the S-SCSE algorithm: i) no tuning (in which $\sigma_{h,k}^2=1,\forall k$); ii) EM (in which the EM algorithm in Section~\ref{Para-tun} is used for learning $\{\sigma_{h,k}^2\}$);  iii) lower bound (in which $\{\sigma_{h,k}^2\}$ are exactly known by the receiver in prior). From Fig.~\ref{EM}, we see that all the three approach perform close to each other when $-10\log_{10}(\sigma_{h,min}^2) = 10$ dB, whereas the EM approach significantly outperforms the no tuning approach and performs close to the lower bound for $-10\log_{10}(\sigma_{h,min}^2) = 15$ and $20$ dB. This implies that the S-SCSE algorithm with EM tuning is able to efficiently handle the effect of large-scale fading.

We now compare the performance of the various schemes in the presence of large-scale fading. In simulations, we set $-10\log_{10}(\sigma_{h,min}^2) = 20$ dB, and the EM algorithm in Section~\ref{Para-tun} is employed for the tuning of $\{\sigma_{h,k}^2\}$. The other settings are the same as those in
Fig.~\ref{X}(a). From Fig.~\ref{Fading20}, we see that the trends of the curves are very similar to those in Fig.~\ref{X}(a), except that the SNR is shifted by about $15$ dB.

Fig.~\ref{throughput} shows the throughput of the JCSE, BD, and S-SCSE with 16-QAM and Gray mapping versus SNR. We say that a system performs successful recovery when
BER $< 10^{-3}$. For the S-SCSE and JCSE schemes, for each given SNR, we increase the number of pilots $T_{\rm P}$ until the system performs successful recovery. For BD, $T_{\rm P}$ is fixed at $3$. Then, the throughput is calculated by $4K(1-T_{\rm P}/T)$ bit per channel use.
The other settings are $K=20$, $N=200$, $T=50$, and $\sigma_{h,min}^2=1$.
For example, in the third subfigure of $\rho=0.3$, when $\text{SNR}=24$ dB, the required pilots to ensure successfully recovery for S-SCSE and JCSE are $T_{\rm P} = 3$ and  $T_{\rm P} = 9$, respectively. So the throughput of the two schemes are calculated by $4\times 20\times(1-3/50)=75.2$ and
$4\times 20\times(1-9/50)=65.6$, respectively.
From Fig.~\ref{throughput}, we see that S-SCSE considerably outperforms the JCSE scheme for $\rho=0.1$, $0.2$, and $0.3$. For $\rho=0.4$, the sparsity level $\rho$ is too large so that the blind matrix factorization cannot provide much useful information. Both the S-SCSE and JCSE schemes rely on the knowledge of pilots for channel-and-signal estimation, and perform closely in Fig.~\ref{throughput}(d).
For comparison, we also include the SNR threshold beyond which the blind detection scheme is able to perform successful recovery.
 Note that the threshold is not included in Fig.~\ref{throughput}(d) since for $\rho=0.4$, the blind detection scheme does not work in the SNR range of interest.
 We see that the blind detection scheme works well only when the SNR is sufficiently high.
This demonstrates the advantage of semi-blind detection.

\section{Conclusions}
\label{sec.Con}
In this paper, we proposed a semi-blind  signal detection scheme for uplink massive MIMO in which short sequences are inserted into user packets and the knowledge of pilots is intergraded into the message passing algorithm for reliable matrix factorization.
We derived two semi-blind estimation algorithms, namely SCSE and S-SCSE, based on the massage-passing principles.
In specific, the S-SCSE is a simplified version of SCSE with much lower computation complexity, but achieve almost the same performance.
We showed that our proposed semi-blind scheme substantially outperforms the existing blind detection and training-based schemes in the short-pilot regime.



\begin{thebibliography}{50}
\providecommand{\url}[1]{#1}
\csname url@samestyle\endcsname
\providecommand{\newblock}{\relax}
\providecommand{\bibinfo}[2]{#2}
\providecommand{\BIBentrySTDinterwordspacing}{\spaceskip=0pt\relax}
\providecommand{\BIBentryALTinterwordstretchfactor}{4}
\providecommand{\BIBentryALTinterwordspacing}{\spaceskip=\fontdimen2\font plus
\BIBentryALTinterwordstretchfactor\fontdimen3\font minus
  \fontdimen4\font\relax}
\providecommand{\BIBforeignlanguage}[2]{{%
\expandafter\ifx\csname l@#1\endcsname\relax
\typeout{** WARNING: IEEEtran.bst: No hyphenation pattern has been}%
\typeout{** loaded for the language `#1'. Using the pattern for}%
\typeout{** the default language instead.}%
\else
\language=\csname l@#1\endcsname
\fi
#2}}
\providecommand{\BIBdecl}{\relax}
\BIBdecl

\bibitem{b1}
E.~G. Larsson, O.~Edfors, F.~Tufvesson, and T.~L. Marzetta, ``Massive {MIMO}
  for next generation wireless systems,'' \emph{IEEE Commun. Mag.}, vol.~52,
  no.~2, pp. 186--195, Feb. 2014.

\bibitem{b2}
T.~L. Marzetta, ``Noncooperative cellular wireless with unlimited numbers of
  base station antennas,'' \emph{IEEE Trans. Wireless Commun.}, vol.~9, no.~11,
  pp. 3590--3600, Nov. 2010.

\bibitem{b3}
H.~Q. Ngo, E.~G. Larsson, and T.~L. Marzetta, ``Energy and spectral efficiency
  of very large multiuser {MIMO} systems,'' \emph{IEEE Trans. Commun.},
  vol.~61, no.~4, pp. 1436--1449, Apr. 2013.

\bibitem{b4}
J.~Hoydis, S.~T. Brink, and M.~Debbah, ``Massive {MIMO} in the {UL/DL} of
  cellular networks: How many antennas do we need?'' \emph{IEEE J. Sel. Areas
  Commun.}, vol.~31, no.~2, pp. 160--171, Feb. 2013.

\bibitem{lu2014overview}
L.~Lu, G.~Y. Li, A.~L. Swindlehurst, A.~Ashikhmin, and R.~Zhang, ``An overview
  of massive {MIMO}: Benefits and challenges,'' \emph{IEEE J. Sel.topics in
  signal process.}, vol.~8, no.~5, pp. 742--758, 2014.

\bibitem{HMT}
T.~L. Marzetta, ``How much training is required for multiuser {MIMO}?''
  \emph{in Fortieth Asilomar Conf. on Signlas Systems, \& Computers}, pp.
  359--363, Oct. 2006.

\bibitem{b5}
M.~Coldrey and P.~Bohlin, ``Training-based {MIMO} systems-part i: performance
  comparison,'' \emph{IEEE Trans. Signal Process.}, vol.~55, no.~11, pp.
  5464--5476, Nov. 2007.

\bibitem{b10}
X.~Yuan, C.~Fan, and Y.~Zhang, ``Fundamental limits of training-based uplink
  multiuser {MIMO} systems,'' \emph{IEEE Trans. Wireless Commun.}, 2018, to
  appear.

\bibitem{ShiJin}
C.~Wen, C.~Wang, S.~Jin, K.~Wong, and P.~Ting, ``Bayes-optimal joint
  channel-and-data estimation for massive {MIMO} with low-precision {ADC}s,''
  \emph{IEEE Trans. Signal Process.}, vol.~64, no.~10, pp. 2541 -- 2556, 2016.

\bibitem{2003much}
B.~Hassibi and B.~M. Hochwald, ``How much training is needed in
  multiple-antenna wireless links?'' \emph{IEEE Trans. Inf. Theory}, vol.~49,
  no.~4, pp. 951--963, 2003.

\bibitem{honig1995blind}
M.~Honig, U.~Madhow, and S.~Verdu, ``Blind adaptive multiuser detection,''
  \emph{IEEE Trans. Inf. Theory}, vol.~41, no.~4, pp. 944--960, 1995.

\bibitem{wang1998blind}
X.~Wang and H.~V. Poor, ``Blind multiuser detection: A subspace approach,''
  \emph{IEEE Trans. Inf. Theory}, vol.~44, no.~2, pp. 677--690, 1998.

\bibitem{zheng2003}
L.~Zheng and D.~N.~C. Tse, ``Diversity and multiplexing: A fundamental tradeoff
  in multiple-antenna channels,'' \emph{IEEE Trans. Inf. Theory}, vol.~49,
  no.~5, pp. 1073--1096, 2003.

\bibitem{S3}
W.~U. Bajwa, J.~Haupt, A.~M. Sayeed, and R.~Nowak, ``Compressed channel
  sensing: A new approach to estimating sparse multipath channels,'' \emph{IEEE
  Trans. Signal Process.}, vol.~98, no.~6, pp. 1058--1076, Jun 2010.

\bibitem{yin2013}
H.~Yin, D.~Gesbert, M.~Filippou, and Y.~Liu, ``A coordinated approach to
  channel estimation in large-scale multiple-antenna systems,'' \emph{IEEE J.
  Sel. Areas in Commun.}, vol.~31, no.~2, pp. 264--273, 2013.

\bibitem{muller2014}
R.~R. M{\"u}ller, L.~Cottatellucci, and M.~Vehkaper{\"a}, ``Blind pilot
  decontamination,'' \emph{IEEE J. Sel. Topics in Signal Process.}, vol.~8,
  no.~5, pp. 773--786, 2014.

\bibitem{masood2015}
M.~Masood, L.~H. Afify, and T.~Y. Al-Naffouri, ``Efficient coordinated recovery
  of sparse channels in massive mimo,'' \emph{IEEE Trans. Signal Process.},
  vol.~63, no.~1, pp. 104--118, 2015.

\bibitem{b15}
J.~Zhang, X.~Yuan, and Y.~Zhang, ``Blind signal detection in massive {MIMO}:
  Exploiting the channel sparsity,'' \emph{IEEE Trans. Commun.}, vol.~66,
  no.~2, pp. 700--712, Feb. 2018.

\bibitem{Philip}
J.~P. Vila and P.~Schniter, ``Expectation-maximization gaussian-mixture
  approximate message passing,'' \emph{IEEE Trans. Signal Process.}, vol.~61,
  no.~19, pp. 4658--4672, 2013.

\bibitem{b16}
J.~T. Parker, P.~Schniter, and V.~Cevher, ``Bilinear generalized approximate
  message passing part {I}: Derivation,'' \emph{IEEE Trans. Signal Process.},
  vol.~62, no.~22, pp. 5839--5853, Nov. 2014.

\bibitem{b17}
D.~L. Donoho, A.~Maleki, and A.~Montanari, ``Message-passing algorithms for
  compressed sensing,'' \emph{Proceedings of the National Academy of Sciences
  of the United States of America}, vol. 106, no.~45, pp. 18\,914--18\,919,
  2010.

\bibitem{liuhang}
H.~Liu, X.~Yuan, and Y.~Zhang, ``Super-resolution blind channel-and-signal
  estimation for massive {MIMO} with arbitrary array geometry,'' \emph{arXiv
  preprint arXiv:1810.01059}.

\bibitem{dempster1977maximum}
A.~P. Dempster, N.~M. Laird, and D.~B. Rubin, ``Maximum-likelihood from
  incomplete data via the {EM} algorithm,'' \emph{J. Roy. Statist. Soc.},
  vol.~39, no.~1, pp. 1--22, 1977.


\bibitem{tropp2007signal}
J.~A. Tropp and A.~C. Gilbert, ``Signal recovery from random measurements via
  orthogonal matching pursuit,'' \emph{IEEE Trans. Inf. Theory}, vol.~53,
  no.~12, pp. 4655--4666, 2007.

\bibitem{needell2009cosamp}
D.~Needell and J.~A. Tropp, ``Cosamp: Iterative signal recovery from incomplete
  and inaccurate samples,'' \emph{Appl. Comput. Harmonic Anal.}, vol.~26,
  no.~3, pp. 301--321, 2009.

\bibitem{chen2018structured}
L.~Chen, A.~Liu, and X.~Yuan, ``Structured turbo compressed sensing for massive
  mimo channel estimation using a markov prior,'' \emph{IEEE Trans. Veh.
  Technol.}, vol.~67, no.~5, pp. 4635--4639, 2018.


\bibitem{CS2008}
E. J. Candes and M. B. Wakin, ``An introduction to compressive sampling,''
 \emph{IEEE Signal Process. Mag.}, vol.~25,
  no. 2, pp. 21¨C30, Mar. 2008.


\bibitem{our}
W.~Yan and X.~Yuan, ``Semi-blind signal detection for uplink massive {MIMO}
  with channel sparsity,'' in {\emph{proc. 2019 IEEE }, {\rm
  Shanghai, China}}.



\end{thebibliography}

\end{document}